\documentclass[
    reprint,superscriptaddress,
    bibnotes,
    amsfonts,amsmath,amssymb,floatfix,longbibliography,aps,prapplied
]{revtex4-2}

\usepackage[dvipdfmx]{graphicx}
\usepackage{amsthm,mathrsfs,dsfont}
\usepackage{mathtools}
\usepackage{braket}
\usepackage{bm}
\usepackage{booktabs}
\usepackage[colorlinks,linkcolor=blue,citecolor=blue]{hyperref}

\newcommand{\tr}{\operatorname{Tr}}
\newcommand{\ham}{\hat{H}}
\newcommand{\s}{\hat{\sigma}}
\DeclareMathOperator*{\argmax}{arg\,max}

\graphicspath{{./fig/}}  
\newcommand\calA{\mathcal{A}}

\newcommand\calI{\mathcal{I}}

\newcommand\calO{\mathcal{O}}

\newcommand\calR{\mathcal{R}}

\newcommand{\ha}{\hat{a}}
\newcommand{\hb}{\hat{b}}
\newcommand{\hn}{\hat{n}}
\newcommand{\hO}{\hat{O}}

\newcommand{\eref}[1]{Eq.~\eqref{#1}}
\newcommand{\tref}[1]{Table~\ref{#1}}
\newcommand{\fref}[1]{Fig.~\ref{#1}}

\newcommand{\sref}[1]{Sec.~\ref{#1}}
\newcommand{\aref}[1]{Appendix~\ref{#1}}
\newcommand{\ccite}[1]{Ref.~\cite{#1}}
\newcommand{\cscite}[1]{Refs.~\cite{#1}}
\allowdisplaybreaks[1]
\usepackage{datetime2}
\begin{document}
% --------------------  TITLE  --------------------
\title{Deep reinforcement learning for preparation of thermal and prethermal quantum states}

% ------------  AUTHORS AND AFFILIATIONS ----------
\author{Shotaro~Z.~Baba}
\email{baba@noneq.t.u-tokyo.ac.jp}
\affiliation{Department of Applied Physics, The University of Tokyo, Hongo, Bunkyo-ku, Tokyo 113-8656, Japan}

\author{Nobuyuki Yoshioka}
\email{nyoshioka@ap.t.u-tokyo.ac.jp}
\affiliation{Department of Applied Physics, The University of Tokyo, Hongo, Bunkyo-ku, Tokyo 113-8656, Japan}
\affiliation{Theoretical Quantum Physics Laboratory, RIKEN Cluster for Pioneering Research (CPR), Wako-shi, Saitama 351-0198, Japan}
\affiliation{JST, PRESTO, 4-1-8 Honcho, Kawaguchi, Saitama, 332-0012, Japan}

\author{Yuto Ashida}
\affiliation{Department of Physics, The University of Tokyo, Hongo, Bunkyo-Ku, Tokyo 113-0033, Japan}

\author{Takahiro Sagawa}
\affiliation{Department of Applied Physics, The University of Tokyo, Hongo, Bunkyo-ku, Tokyo 113-8656, Japan}
\affiliation{Quantum-Phase Electronics Center (QPEC), The University of Tokyo, Tokyo 113-8656, Japan}
%\email{email}
%\date{\DTMnow}
%\date{\today}
% --------------------  ABSTRACT  --------------------
\begin{abstract}
    %We propose the strategy to prepare thermal quantum many-body states by reinforcement learning with the reward leveraging a few local observables.
    We propose a method based on deep reinforcement learning that efficiently prepares a quantum many-body pure state in thermal or prethermal equilibrium.
    %Our strategy does not require fidelity, which is too expensive to measure in quantum many-body systems from a practical point of view.
    The main physical intuition underlying the method is that the information on the equilibrium states can be efficiently encoded/extracted by focusing on only a few local observables, relying on the typicality of equilibrium states. 
    Instead of resorting to the expensive preparation protocol that adopts global features such as the quantum state fidelity, we show that the equilibrium states can be efficiently prepared only by learning the expectation values of local observables.
    We demonstrate our method by preparing two illustrative examples: Gibbs ensembles in non-integrable systems and generalized Gibbs ensembles in integrable systems.    
    %We apply this strategy to Gibbs ensembles in non-integrable systems and generalized Gibbs ensembles in integrable systems.
    Pure states prepared solely from local observables are numerically shown to successfully encode the macroscopic properties of the equilibrium states.
    %, thanks to the typicality of the pure states of quantum many-body states.
    Furthermore, we find that the preparation errors, with respect to the system size, decay exponentially for Gibbs ensembles and polynomially for generalized Gibbs ensembles,
    which are in agreement with the finite-size fluctuation within thermodynamic ensembles.
    Our method paves a path toward studying the thermodynamic and statistical properties of quantum many-body systems in quantum hardware.
    %and we would achieve more efficient preparations in the larger systems.
    %These futures may lead to the application of our method to Hamiltonian learning in large experimental many-body systems.
\end{abstract}
\maketitle
\section{Introduction}
    %\bout{
    %The recent advances in the experimental platform of quantum many-body simulations, 
    %such as ultracold atoms~\cite{bloch_many-body_2008,bloch_quantum_2012,ueda_quantum_2020}, 
    %trapped ions~\cite{blatt_quantum_2012}, 
    %and superconducting qubits~\cite{xu_emulating_2018,yan_strongly_2019,neill_ergodic_2016}, 
    %attract much attention to the development of the technology of manipulating quantum many-body states.}
    %More specifically, 
    The preparation of a desired quantum many-body state is an essential task that plays a significant role in quantum computing~\cite{nielsen_quantum_2010,ladd_quantum_2010}, quantum metrology~\cite{giovannetti_advances_2011}, and quantum communication~\cite{gisin_quantum_2002}.
     %and elucidating novel material properties.
    %Considering the practical application of the quantum-state preparation, the realistic experimental strategy to design the protocols for manipulating the dynamics of quantum states.
    Specifically, the thermal state is one of the most important targets in quantum state preparation tasks~\cite{terhal_problem_2000,poulin_sampling_2009,sugiura_canonical_2013,brandao_finite_2019,wu_variational_2019,chowdhury_variational_2020,wang_variational_2021} from both theoretical and experimental viewpoints.
    A common strategy is to employ numerical methods such as CRAB~\cite{doria_optimal_2011,caneva_chopped_2011}, GRAPE~\cite{khaneja_optimal_2005}, and Krotov~\cite{krotov_global_1996}. 
    However, it must be noted that all these methods suffer from the exponential growth of computational cost and, furthermore, require detailed knowledge about the nonequilibrium properties of the system. 
    Therefore, it is desirable to construct a preparation protocol that employs only a little knowledge during the learning task.

    %The thermal state of a given Hamiltonian is essential for many quantum algorithms such as quantum simulated annealing~\cite{somma_quantum_2008}, quantum machine learning~\cite{biamonte_quantum_2017,kieferova_tomography_2017}, and quantum semi-definite programming~\cite{brandao_quantum_2017}.
    %On the other hand, the thermal or stationary state of an unknown Hamiltonian can be used for the estimation of the unknown parameter of the Hamiltonian, called Hamiltonian learning~\cite{rudinger_compressed_2015,bairey_learning_2019,anshu_sample-efficient_2021}.
    %In most cases of constructing experimental preparation policies in general many-body systems, 
    %due to the lack of our theoretical understanding of nonequilibrium dynamics in such systems, 
    %numerical optimization using numerical simulations, such as CRAB~\cite{doria_optimal_2011,caneva_chopped_2011}, GRAPE~\cite{khaneja_optimal_2005}, and Krotov~\cite{krotov_global_1996}, are one essential method~\cite{glaser_training_2015}.
    %Nevertheless, numerical simulations require detailed knowledge about the system. %#, 
    %To take advantage of the Hamiltonian learning from stationary states, we need to develop the method to prepare the stationary states of an unknown Hamiltonian using a little knowledge about the system.
    %and the exponential growth of the Hilbert space dimension of quantum many-body systems makes numerical simulations hard.

    %\blue{(NY: Error in citation comes from \{yoshioka learning 2018\}, which is included in library.bib)}
    A surging technology to extract the essential features in quantum systems with a prohibitively large exploration space is machine learning, which has exemplified its capacity across a wide range of physics~\cite{mehta_high-bias_2019,carleo_machine_2019, yoshioka_transforming_2019,nomura_purifying_2021,glasser_neural-network_2018,gao_efficient_2017,lu_efficient_2019,hibat-allah_recurrent_2020,sharir_deep_2020,pfau_ab-initio_2020,carleo_solving_2017,lagaris_artificial_1997, yoshioka_constructing_2019,vicentini_variational_2019,hartmann_neural-network_2019, carrasquilla_reconstructing_2019,torlai_neural-network_2018,torlai_latent_2018, carrasquilla_machine_2017, yoshioka_learning_2018,erdman_identifying_2022,erdman_driving_2022}. 
    Successful applications include representations of quantum many-body states with neural networks (NNs)~\cite{yoshioka_transforming_2019,nomura_purifying_2021,yoshioka_solving_2021, glasser_neural-network_2018,gao_efficient_2017,lu_efficient_2019,hibat-allah_recurrent_2020,sharir_deep_2020,pfau_ab-initio_2020,carleo_solving_2017,lagaris_artificial_1997, yoshioka_constructing_2019,vicentini_variational_2019,hartmann_neural-network_2019,nagy_variational_2019}, 
    quantum state tomography~\cite{carrasquilla_reconstructing_2019,torlai_neural-network_2018,torlai_latent_2018}, 
    and phase classification~\cite{carrasquilla_machine_2017, yoshioka_learning_2018,tibaldi_unsupervised_2022}, to name a few.
    %making use of its capability of finding the solution from the exponentially large exploration spaces effectively.
    In particular, a branch of machine learning called reinforcement learning (RL)~\cite{sutton_reinforcement_2018} has been recognized as a powerful tool to perform quantum state preparation~\cite{bukov_reinforcement_2018,bukov_reinforcement_2018-1,yao_reinforcement_2021,sivak_model-free_2022,wang_deep_2020,borah_measurement-based_2021,niu_universal_2019,ding_breaking_2021,an_quantum_2021}.
    RL is designed to discover an efficient policy that maximizes a given reward through trial-and-error learning on the behavior of the environment.
    Several previous studies utilize the algorithm that adopts the deep RL framework; the quantitative evaluation of the action, or the reward, determined by the algorithm makes full use of the capability of NNs to approximate high-dimensional nonlinear functions~\cite{an_quantum_2021,sgroi_reinforcement_2021,niu_universal_2019,borah_measurement-based_2021,wang_deep_2020,sivak_model-free_2022,ding_breaking_2021,fosel_reinforcement_2018}.
    While the bulk of the previous works choose fidelity as the reward, its computation for quantum many-body systems requires exponentially large resources in either numerics or experiment, and thus fidelity is not practical to scale up.
    %measuring fidelity in realistic experiments requires a large sample size using, e.g., quantum state tomography~\cite{flammia_direct_2011}.
    %This fact makes using fidelity as a reward too expensive to prepare states in many-body systems from a practical point of view.

    %Therefore, the appropriate reward function for the target state instead of fidelity is desirable.

    In this work, we propose a deep-RL-based method that only relies on local measurements to prepare thermal and prethermal pure states described by Gibbs and generalized Gibbs ensembles (GGEs)~\cite{rigol_thermalization_2008,rigol_relaxation_2007,vidmar_generalized_2016,dalessio_quantum_2016}.
    The underlying physical intuition is that we may take advantage of the typicality of equilibrium states~\cite{goldstein_canonical_2006,popescu_entanglement_2006,sugiura_canonical_2013,sugiura_thermal_2012} to prepare them solely using local observables, without reliance on global features such as fidelity.
    %the limited knowledge 
    %knowledge only of the local property of the target system to prepare equilibrium states such as Gibbs ensembles and generalized Gibbs ensembles.
    %We design the new appropriate reward function of RL leveraging the typicality of quantum many-body states~\cite{goldstein_canonical_2006,goldstein_normal_2010,tasaki_typicality_2016} and do not use fidelity in the RL training.
    %In addition, we perform the finite-size scaling of the accuracy of the preparation with the system size.
    Numerically, we find that although the deep RL agent is only informed of the local information on the thermodynamic ensembles, the accuracy of the prepared state improves exponentially with the system size for Gibbs ensembles, whereas the improvement is polynomial for GGEs.
    %We furthermore emphasize that, for the case of Gibbs ensembles, such a scaling is observed not only within local observables used for training, but also for observables with larger locality.
    %Efficient preparation scheme found by the proposed method may open a path to elucidate deeper understanding on exotic quantum many-body systems in equilibrium, e.g., via Hamiltonian learning~\cite{rudinger_compressed_2015,bairey_learning_2019,anshu_sample-efficient_2021}.
    %of unknown Hamiltonian corresponding
    %Such features may help our method be applied to Hamiltonian learning~\cite{rudinger_compressed_2015,bairey_learning_2019,anshu_sample-efficient_2021}, 
    %where the thermal or stationary state of an unknown Hamiltonian is used for the estimation of the unknown parameter of the Hamiltonian, 
    %in large experimental systems.
    %\blue{NY: Maybe mention Hamiltonian learning here? maybe also at Discussion as well}
    
    The remainder of the paper is organized as follows. 
    In Sec.~\ref{sec:general_RL}, we present an overview of the framework of the deep RL. 
    The application of RL to quantum state preparation is described in Sec.~\ref{sec:rl_preparation}, which includes the core proposal of our work, i.e., the local preparation of thermal and prethermal pure quantum states leveraging the typicality of equilibrium states. 
    After the framework is presented,  we give the numerical demonstration for preparation of equilibrium states described by Gibbs ensembles and GGEs in Sec.~\ref{sss:ge} and~\ref{sss:gge}, respectively.
    Finally, we give our conclusions and a discussion in Sec.~\ref{sss:conclusion}.
    %In Sec.~\ref{sss:method}, we introduce the framework of RL, key concepts, and our method of quantum state preparation.
    %In \sref{sss:ge}, we provide the demonstration of our method in a Gibbs ensemble.
    %In \sref{sss:gge}, we describe the result of our method in a generalized Gibbs ensemble. % as a more intriguing example.
    %In \sref{sss:conclusion}, we conclude our work.
%\section{Method}\label{sss:method}
%\sec
    \section{Reinforcement learning}\label{sec:general_RL}
\begin{figure}[tbp]
            \centering
            \includegraphics[width=\linewidth]{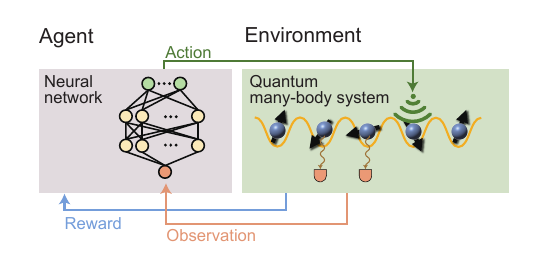}
            \caption{
            A graphical illustration of the deep RL framework to prepare thermal and prethermal pure quantum states for thermodynamic ensembles.
            The agent represented by the deep NN is trained to discover a policy that maximizes the cumulative reward, which is computed based on the local observables of the quantum many-body system (environment).
            At each time step, the NN takes the result of observation as input and outputs the action-value function, from which the action (or the control operation) on the environment is determined.
            In this work, the action is chosen from a set of unitaries that are presumably available in the preparation protocol.
            %The deep RL setting of this work.
            %By interacting with environments, the agent discovers optimal policies to maximize the given reward by trial-and-error learning.
            %The dynamics of the quantum many-body systems (environment) are manipulated by the controller (agent) with the neural network (NN).
                %The NN receives the observation as its input from the environment and output the action-value function, which tells the expected future cumulative reward for each given action.
                %According to this output, the agent chooses the next action.
                %Using the reward, the agent updates the parameters of the NN so that the NN appropriates the optimal action-value function.
                %In this work, the action set is the set of the unitaries, and
                %the controller looks for the best unitary sequence to maximize the given reward, which tells the distance between the state in preparation \(\rho_{t}\) and the target state \(\rho_{\mathrm{target}}\).
            }\label{fig:RL_framework}
        \end{figure}

        The general framework of the RL can be concisely expressed as a procedure to train an agent how to interact with the environment through the optimization of cumulative reward~[see Fig.~\ref{fig:RL_framework}].
        The strengths of deep RL, which are why we choose to employ it, are the high degree of freedom in reward design that makes the algorithm independent of the actual model and its ability to handle a huge search space of total actions, which amounts to \(15^{200}\approx 10^{235}\) in our demonstration in integrable systems in \sref{sss:gge}.
        The corresponding quantities (game tree complexity) of chess, shogi (Japanese chess), and Go, which are canonical environments for high-performance planning, are \(\approx 10^{123}\), \(10^{226}\) and \(10^{360}\), respectively~\cite{Allis1994-mk,Iida2002-zj,Silver2016-lv}.
        The search space size of our problem is larger than that of shogi and deep RL is a reasonable choice to achieve higher performance~\footnote{The authors of \ccite{Silver2018-di} applied deep RL to defeat a world champion program of shogi, which is based on a highly optimized alpha-beta search engine with many domain-specific adaptations.}. 
        
        % \red{temp}
        % As an optimization method for the state preparation protocol, we employed deep RL in this study.
        % DeepMind Technologies, a subsidiary of Google, applied deep RL to defeat a world champion program of shogi, which is based on a highly optimized alpha-beta search engine with many domain-specific adaptations~[Silver, D. \textit{et al.}, Science \textbf{362}, 1140 (2018).].
        
        The goal of  RL is to discover the best policy \(\pi\) that outputs the sequence of actions \(\calA = \{a_t\}_t\) based on observations of the environment \(\calO = \{o_t\}_t\), such that the feedback realizes the most desired behavior quantified by the rewards \(\calR = \{r_t\}_t\).
        Typically, all the events are discretized so that each value can be well-defined at each time step \(t\). 
        
        A practical strategy widely used in the community of machine learning is Q-learning. 
        Here, one aims to find the best approximation of the optimal action-value function as~\cite{sutton_reinforcement_2018}
        \begin{align}
            Q^{*}(o,a)=\max_{\pi}\mathbb{E}_{\pi}\left\lbrack r_{t}+\sum_{n=1}^{\infty}\gamma^{n} r_{t+n} \middle|o_{t}=o,\ a_{t}=a,\pi \right\rbrack ,\label{eq:Q}
        \end{align}        
        which is the maximum sum of rewards \(r_{t}\) discounted by \(\gamma\) (\(0<\gamma<1\)) in a stochastic policy that chooses action according to some probability distribution as \(\pi\left\lparen  a|o \right\rparen=\Pr\left\lparen  a|o \right\rparen\).
        A powerful flavor of Q-learning uses the deep NNs to represent the action-value function, and hence referred to as the deep RL algorithms~\cite{li_deep_2018,henderson_deep_2019}. 
        The extraordinary representative power of NNs has been found to facilitate successful applications of deep RL algorithms in numerous fields that are not necessarily limited to computer science but also include the natural science, the materials sciences, and so on. %~\cite{??} \blue{NY: cite something? this is optional}.
        %Using the batches of stored results, we update the parameters of the NN\@.
        %After optimizing the NN, we can obtain the optimal deterministic policy \(\pi^{*}(a|o)=1\) only if \(a=\argmax_{a'}Q(o,a')\).

        In this paper, we focus on a non-distributed
        implementation~\cite{stooke_rlpyt_2019} of a deep RL algorithm called R2D2~\cite{kapturowski_recurrent_2019}.
        R2D2 is a type of deep Q-learning algorithm~\cite{mnih_human-level_2015} 
        %, a family of deep RL~\cite{li_deep_2018,henderson_deep_2019},
        and assumes that the agent can obtain partial information about the state of the environment.
        As we show the architecture in Appendix~\ref{app:layout}, the NN used in R2D2 includes a Long Short-Term Memory (LSTM) layer, and therefore the action-value function \(Q\) computed by the NN at step \(t\) depends not only on the instant observation \(o_{t}\) 
        but also on the previous observations \({\left\lbrace o_{t'} \right\rbrace}_{t' \leq t}\)~\cite{hochreiter_long_1997}.
        %Thus, strictly speaking, the action-value function should be written as \(Q({\left\lbrace o_{s} \right\rbrace_{s=0,\ldots ,t}},a)\). 
        This feature enables the NN to handle time-series inputs and develop the capability in a partially observed environment.

    \section{Reinforcement learning for quantum state preparation}
    \label{sec:rl_preparation}
    \subsection{Global state preparation}
    \label{subsec:global_prep}
        Next, we review the general protocol to prepare a desired isolated quantum many-body state using the deep RL framework.
        Specifically, we aim to prepare a target quantum state \(\ket{\psi_{\rm target}}\) from an easily prepared quantum state \(\ket{\psi_0}\), assuming that a set of unitary \(\{U_i\}_i\) is available at any time step. By finding the best sequence of unitaries, we try to approximate the target state as
        \begin{equation}
            \ket{\psi_{\rm target}} \approx \prod_t U_{i_t} \ket{\psi_0}.
        \end{equation}
        It is straightforward to see that such a problem setup is in a great connection with the RL\@; we identify the quantum many-body system with the environment and the available set of unitaries \(\{U_i\}_i\) with the action candidates \(\{a_t\}\) at each time step.
        
        Regarding the observation \(o_t\), many works have proposed to use the results for the measurements on the target system~\cite{sivak_model-free_2022,wang_deep_2020,borah_measurement-based_2021}. 
        Meanwhile, when both the initial and target states are fixed during the whole training, we can expect that the action history \(\{a_{t'}\}_{t'\leq t}\) will contain enough information to find out the desired protocol ~\cite{bukov_reinforcement_2018-1,yao_reinforcement_2021}. %(similar choice also found in Refs.~\cite{bukov_reinforcement_2018-1,yao_reinforcement_2021}). 
        %As we mention in the next section, we take the same strategy in our work.

        As for the reward \(r_t\), numerous existing works have considered global features such as the fidelity \(F(\rho_{t},\rho_{\mathrm{target}})={\left\lparen  \tr \sqrt{\sqrt{\rho_{t}}\rho_{\mathrm{target}}\sqrt{\rho_{t}}} \right\rparen}^{2}\)~\cite{bukov_reinforcement_2018,bukov_reinforcement_2018-1,niu_universal_2019,an_quantum_2021},
        where \(\rho_t\) is the density operator of the controlled system at time step \(t\).
        Hereafter, we refer to such protocols as {\it global} state preparation protocols. 
        These methods have been used to successfully prepare ground states of quantum many-body spin systems~\cite{bukov_reinforcement_2018}, metastable states of the quantum Kapitza oscillator~\cite{bukov_reinforcement_2018-1}, and highest excited states of multi-level quantum systems~\cite{an_quantum_2021}.
        %, while the cost to obtain such quantities becomes practically prohibitive as the system size becomes larger.
        
        %on the quantum state as observation \(o_t\).
    
        Having bridged between the notions in quantum control and RL, we can train the NN-based agent to find the best  control on the quantum system via searching for an approximation of 
        the optimal action-value function \(Q^{*}(o,a)\)~\eqref{eq:Q}.
        Note that the evolution of the quantum state may either be experimentally implemented or numerically simulated, as long as the reward function for the deep RL agent can be readily obtained.
        Once the training is completed, we determine the preparation protocol as \(\{a^*_t\}_t\) by choosing the actions so that \(a^*_{t}=\argmax_{a}Q(o_{t},a)\) at each time step \(t\), which maximizes the reward.
        
        %\blue{(NY:The following could be included in the next subsection?)}
        
    \subsection{Local state preparation}
    \label{subsec:local_prep}
        We are now ready to describe the preparation protocol for thermal and prethermal pure quantum states that relies solely on local observables, instead of querying for costly global features such as the fidelity.
        Below, as opposed to {\it global} state preparation, we refer to the following scheme as {\it local} state preparation.
        
        The central idea of local state preparation is to leverage the typicality of pure quantum states~\cite{goldstein_canonical_2006,popescu_entanglement_2006,sugiura_canonical_2013,sugiura_thermal_2012}.
        Typicality refers to the fact that the overwhelming majority of pure quantum states with the same local conserved quantities are indistinguishable means of local observables. Such states are also deemed to be {\it macroscopically indistinguishable}. 
        %\blue{NY: Please define local conserved quantities, which is ambiguous in the present manuscript}
        %We may alternatively express that the average difference between the reduced density operators of the states with the same local conserved quantities becomes exponentially small with respect to their system size.
        %coincide for large systems. % in the thermodynamic limit.
        %This implies that w
        It is natural to expect that, by utilizing the typicality, we can prepare a state that encodes the macroscopic properties of the equilibrium states solely by controlling the local observables.
        Specifically, we perform local preparation by making the expectation values of the local observables close to the equilibrium state and then letting the system relax to equilibrium through a unitary evolution, without any control.

        We may utilize various forms of reward to learn the local observables of the target ensemble.
        %To make the expectation values of local observables close to those of the target states,
        In this work, we formulate the reward function as the inverse of the deviations of the expectation values of the local observables from that of the target states:
        \begin{align}
            r_{t}\coloneqq\frac{1}{\left\lvert M_{t}-M_{\mathrm{target}}\right\rvert+\varepsilon},\label{eq:inv_reward}
        \end{align}
        where 
        \(M_{t}=( \braket{O_{1}}_{t},\braket{O_{2}}_{t},\ldots )\)
        and 
        \(M_{\mathrm{target}}=( \braket{O_{1}}_{\mathrm{target}},\braket{O_{2}}_{\mathrm{target}},\ldots )\)
        are vectors consisting of expectation values  \(\braket{O}_{t({\rm target})} := {\rm Tr}[\rho_{t({\rm target})} O]\). 
        The small constant \(\varepsilon\) is also introduced to prevent divergence of the reward function.

        %\(M\) is the array of the expectation values of the observables we focus on, \(M_{t}=( \braket{O_{1}}_{t},\braket{O_{2}}_{t},\ldots )\), 
        %\(M_{\mathrm{target}}=( \braket{O_{1}}_{\mathrm{target}},\braket{O_{2}}_{\mathrm{target}},\ldots )\), and \(\varepsilon\) is the small constant to prevent the reward from divergence.
        %This reward leads the agent to learn how to reduce the deviation of the expectation value from the target value.
        
        %\blue{NY: please refine this part. Is this necessary in the main text? }
        %As the observation at time step \(t\), we choose the action history, which is the input to the NN\@.
        %The strategy to adopt the action history as the observation is also used in \cscite{bukov_reinforcement_2018-1,yao_reinforcement_2021}.
        %The architecture of the NN, hyperparameters and computational resources used in this work are described in \aref{app:layout}. 

        As a concrete target for the demonstration of our local preparation protocol, we choose Gibbs ensembles and GGEs as illustrative examples.
        The evolution of the quantum many-body state is numerically simulated in an exact manner, while in principle we may also employ approximate methods that rely on, e.g., a variational representation such as tensor networks or neural networks. 
        Alternatively, one may implement the proposed protocol directly on an experimental device as well.
        In the following, we proceed to describe the detailed properties of the thermodynamic ensembles and the expected preparation efficiency in the presence of typicality.

        \subsubsection{Gibbs ensemble for non-integrable systems}
        \label{ss2:describing_local_preparation_GE}
            %Pure states belonging to a given shell of local conserved quantities are typical; they share macroscopic properties that are described by a statistical ensemble. 
            
            Let us consider a non-integrable system with the energy being the only local conserved quantity such  that its equilibrium is described by a Gibbs ensemble. 
            On a related point, pure states belonging to a given microcanonical shell of the system share their macroscopic properties. This class of typicality is referred to as canonical typicality.
            One of the most prominent examples of canonical typicality can be illustrated by the Haar measure on the space of pure states under some constraint \(R\) (e.g., energy) as~\cite{popescu_entanglement_2006}
            \begin{align}
                \braket{\left\lVert\rho_{A}-\Omega_{A}\right\rVert_{1}}\leq \sqrt{\frac{d_{A}^{2}}{d_{R}}},\label{eq:Typicality_scaling}
            \end{align}
            where 
            \(\rho_{A}=\tr_{\bar{A}} \left[ \ket{\psi}\bra{\psi} \right]\) 
            is the reduced density operator obtained by tracing out the 
            the complement of subsystem \(A\) for \(\ket{\psi}\bra{\psi}\)
            with Hilbert space dimension \(d_A\).
            On the other hand,
            \(\Omega_{A}=\tr_{\bar{A}} \left[\mathds{1}_{R}/d_{R}\right]\) is the reduced density operator for the projection operator \(\mathds{1}_R\), i.e., the maximally mixed state in the Hilbert subspace under the constraint \(R\) whose dimension is \(d_R\).
            Note that the bracket \(\braket{\cdot}\) concerns the average with regard to the Haar measure on the constrained Hilbert space,
            and
            \(\left\lVert A\right\rVert_{1}=\tr \sqrt{A^{\dagger}A}\).
            
            %, so that states in a given microcanonical energy shell have typical macroscopic properties.
            %More specifically, when the system has nothing but energy as local conserved quantities (e.g., non-integrable systems), 
            %Pure states belonging to a given energy shell are typical; they share macroscopic properties that are described by a statistical ensemble. 
            %In particular for called the Gibbs ensemble.
            %The Gibbs ensemble is a tool to elicit these typical macroscopic properties. %The Gibbs ensemble is a tool to characterize such an ensemble.
            %The typicality for the corresponding pure quantum states is called the canonical typicality, which is proved for large systems with regard to the Haar measure on the space of pure states under some constraint \(R\) (e.g., energy) as~\cite{popescu_entanglement_2006}
            %\begin{align}
                %\braket{\left\lVert\rho_{A}-\Omega_{A}\right\rVert_{1}}\leq \sqrt{\frac{d_{A}^{2}}{d_{R}}},\label{eq:Typicality_scaling}
            %\end{align}
            %where \(\rho_{A}=\tr_{\bar{A}} \left[ \ket{\psi}\bra{\psi} \right]\) is the truncation of the complement of subsystem \(A\) for \(\ket{\psi}\bra{\psi}\),
            %\(d_{A}\) is the dimension of the Hilbert space of the subsystem \(A\),
            %\(d_{R}\) is the dimension of the Hilbert space under the constraint \(R\),
            %\(\Omega_{A}=\tr_{\bar{A}} \left[\mathds{1}_{R}/d_{R}\right]\) is the truncation of the maximally mixed state in the Hilbert space under the constraint \(R\),
            %the bracket \(\braket{\cdot}\) means the average regarding the Haar measure on the constrained Hilbert space,
            %and \(\left\lVert A\right\rVert_{1}=\tr \sqrt{A^{\dagger}A}\).
            
            Equation~\eqref{eq:Typicality_scaling} means that the average distance between a randomly chosen pure state and the maximally mixed state in the constrained Hilbert space decays polynomially with the Hilbert space size as \(d_{R}^{-0.5}\), that is, exponentially with the system size in general quantum systems.
            The indistinguishability of the pure quantum state from the microcanonical ensemble leads us to expect that, once a state is prepared to be within the target energy shell, the prepared state captures the macroscopic properties with an accuracy that improves exponentially with the system size~\footnote{
            Gibbs ensembles, which is our target, differs from microcanonical ensembles. Nevertheless, a microcanonical ensemble is locally equivalent to a Gibbs (canonical) ensemble for translation invariant short-ranged Hamiltonians, even if the system is finite~\cite{brandao_equivalence_2015,tasaki_local_2018}.
            In other words, the microcanonical and the canonical expectation values almost coincide when we look at the subsystem that is not too large.}.
            %The maximally mixed state is a microcanonical ensemble when the constraint is the energy.
            %Furthermore, the microcanonical ensemble is locally equivalent to the Gibbs (canonical) ensemble for translation invariant short-ranged Hamiltonians, even if the system is finite~\cite{brandao_equivalence_2015,tasaki_local_2018}.
            %In other words, the microcanonical and the canonical expectation values almost coincide when we look at the subsystem that is not too large.
            %Therefore, based on the scaling of the distance~\eqref{eq:Typicality_scaling}, the accuracy of the local preparation using typicality can be expected to grow exponentially with the system size. 
            
            %Here, we mention the quantum fluctuation regarding the energy.

            We emphasize that the key of local preparation protocol is to encode the prepared state into the target energy shell, which requires more than merely learning the expectation value of the energy.
            %However, merely learning the expectation value of energy does not necessarily assure the prepared pure state to be included in the single energy shell. 
            In other words, even if the expectation value is correctly learned, the prepared state may correspond to a superposition of pure states that belong to other energy shells. % and result in a large fluctuation of macroscopic observables.
            Such a situation may cause deviation in other physical observables.
            In this work, we attempt to address this problem by letting the RL agent learn other macroscopic observables in addition to energy. 
            It is in fact highly nontrivial to determine
            %While it depends on the details of the target system 
            how many observables we need in order to embed the prepared state into the energy shell. We find that
            for the non-integrable transverse-field Ising chain, it suffices to take only the total magnetization \(\sum_l \sigma_l^z/L\) (for the numerical demonstration, see Sec.~\ref{sse:ge_result}).

            %To deal with this problem, we add additional local observables to the reward other than the energy.
            %which may 
            %A pure quantum state does not necessarily belong to a single set of macroscopically indistinguishable states, i.e., a single energy shell.
            %This can be a cause of the failure of the local state preparation.
            %To deal with this problem, we add additional local observables to the reward other than the energy.
            %By making some local observable values close to the target macroscopic quantities, we expect to be able to increase the probability that the preparing pure state belongs to a single macroscopic state.
            %By bringing some local observables close to the target quantities, we expect to increase the probability that the prepared pure state belongs to a single energy shell.
            %As another solution to handle the quantum fluctuation, 

            As another possibly effective method to assure that the prepared state is encoded in the desired energy shell, we propose to incorporate the variance of observables \(\braket{O^2}_{\rm target} - \braket{O}_{\rm target}^2\) into the reward.
            For instance, it is obvious that the energy variance is suppressed when the prepared state is in the target energy shell.
            This is actually not limited to the energy; due to the typicality, we may employ any macroscopic observable for this purpose.

            %we may use quantum measurement results in the reward function instead of the expectation values
            %since quantum measurement results fluctuate according to the quantum fluctuation regarding the corresponding observable.
            %If we design the reward with the deviation of the measurement results from the target macroscopic value, 
            %we expect the agent to learn the protocol that suppresses the fluctuation, and the prepared state belongs to a single energy shell.
        \subsubsection{Generalized Gibbs ensemble for integrable systems}\label{ss2:gge}
            As opposed to the non-integrable systems, integrable systems have an extensive number of conserved quantities, which are also called integrals of motion (IOM) in the literature.
            Rather, the equilibrium states in integrable systems are known to be described by GGEs~\cite{rigol_relaxation_2007}:
            \begin{align}
                \rho_{\mathrm{GGE}}=\frac{\exp\left\lbrack -\sum_{m}\lambda_{m}\hat{\calI}_{m} \right\rbrack }{\tr \left\lbrack \exp\left\lbrack -\sum_{m}\lambda_{m}\hat{\calI}_{m} \right\rbrack  \right\rbrack },\label{eq:gge_definition}
            \end{align}
            where \(\lbrace \hat{\calI}_{m} \rbrace\) is the full set of the IOMs,
            and \(\left\lbrace \lambda_{m} \right\rbrace\) is the corresponding set of the Lagrange multipliers 
            which dictates the distribution over the expectation values of the IOMs\@.

            To discuss the typicality in the set of pure states with close expectation values of IOM, %whose expectation values of integrals of motion are close to a certain value,
            the authors of \ccite{cassidy_generalized_2011} have introduced the notion of a statistical ensemble called the generalized microcanonical ensemble (GME).
            In parallel to the ordinary microcanonical ensemble, the GME is constructed by assigning equal weights to all eigenstates whose IOMs are close to some certain values that identify the ensemble.
            It has been pointed out in \ccite{cassidy_generalized_2011} that the standard deviations of the local observables within such ``a window of IOM" %\(\sigma_{\mathrm{den}}\) in the GME
             decay polynomially as 
            \begin{align}
                \sigma_{\mathrm{\rm loc}}\propto L^{-0.5},\label{eq:gme_decay}
            \end{align}
            where \(L\) is the system size.
            Therefore, by following a parallel discussion as in the case for Gibbs ensembles in non-integrable systems, we may also expect that the local preparation protocol will work for GGEs in integrable systems as well, with its accuracy improving polynomially with the system size.

            Let us remark on another supporting argument based on the truncation of GGEs itself.
            While Eq.~\eqref{eq:gge_definition} takes all possible IOMs into account, we expect that the macroscopic behavior in terms of local observables can be extracted by considering local conserved quantities. 
            As such, here, we aim to capture the truncated alternative of the statistical ensemble by focusing on the {\it local} integrals of motion (LIOM). 
            Specifically, we denote the LIOMs that act at most \(n+1\) neighboring sites as \(\hat{I}_{n}^{\sigma}\), with 
            \(\sigma\) denoting some additional label,
            and we introduce a locality-constrained variant of GGE that is known as truncated GGE (tGGE)~\cite{fagotti_reduced_2013}:
            \begin{align}
                \rho_{\mathrm{tGGE},n_{\mathrm{local}}}=
                \frac{\exp\left\lbrack -\sum_{n=0}^{n_{\mathrm{local}}}
                \sum_\sigma\lambda_{n}^{\sigma}\hat{I}_{n}^{\sigma} \right\rbrack }{\tr \left\lbrack \exp\left\lbrack -\sum_{n=0}^{n_{\mathrm{local}}}
                \sum_\sigma
                \lambda_{n}^{\sigma}\hat{I}_{n}^{\sigma} \right\rbrack  \right\rbrack },
            \end{align}            
            which only includes the LIOMs with \(n\leq n_{\mathrm{local}}\).
            It is natural to expect that a tGGE will give a good approximation of \(\rho_{\rm GGE}\) in terms of local quantities and, furthermore, that \(\tr_{\bar{A}}\left\lbrack \rho_{\mathrm{tGGE},n_{\mathrm{local}}}\right\rbrack\approx \tr_{\bar{A}}\left\lbrack \rho_{\mathrm{GGE}} \right\rbrack\).
            For instance, in Ref.~\cite{fagotti_reduced_2013}, the transverse field Ising chain has been investigated in the integrable regime and it has been found that tGGEs approximate the corresponding GGEs when \(n_{\rm local}\) is larger than the size of subsystem \(A\).
            %their subsystem sizes. %the difference between the tGGE and GGE reduces exponentially with respect to \(n_{\rm local}\). 

            We remark that the local preparation protocol for integrable systems aims to construct a tGGE rather than the original GGE\@. 
            In this sense, we expect that the validity is not assured for observables with higher \(n_{\rm local}\), for which the discrepancy between the GGE and the tGGE is non-negligible. 
            We discuss this point more in detail in Sec.~\ref{sse:gge_result}.
            %,
            %while we find that their distance is smaller than the training error in RL in the system size we focus on in the following.

            %in terms of the following %distance function:
            %\begin{align}
                %\tr_{\bar{A}}\left\lbrack \rho_{\mathrm{tGGE},n_{\mathrm{local}}}\right\rbrack\approx \tr_{\bar{A}}\left\lbrack \rho_{\mathrm{GGE}} \right\rbrack .\label{eq:local_eq_tGGE}
            %\end{align}         
            %}

            %Considering \eref{eq:local_eq_tGGE}, we expect the macroscopic preparation of GGE can be achieved by controlling only LIOMs and letting the systems relax.
            %Furthermore, the accuracy of the local preparation can be expected to grow polynomially with the system size.

\section{Application to Gibbs ensembles}\label{sss:ge}
    \subsection{Model and setup}
        %To begin with, we show the result of the local preparation of the Gibbs ensemble.
        As a demonstration of the local preparation of thermal pure states described by Gibbs ensembles, we consider the transverse field Ising model on a chain with the periodic boundary condition:
        \begin{align}
            \ham_{\mathrm{Ising}}=\sum_{l=1}^{L}\left\lbrack  J\s_l^z\s_{l+1}^z+h\s_{l}^{z}+g\s_{l}^{x} \right\rbrack ,\label{eq:ising}
        \end{align}
        where \(\s_l^x,\ \s_l^y,\ \s_l^z\) are the Pauli operators at site \(l\), \(L\) is the system size, \(J\) is the amplitude of the Ising interaction, and \(h~(g)\) is the strength of the longitudinal~(transverse) magnetic field. 
        In the following, the parameters are fixed as \(J=1,\ h=0.8090,\ g = 0.9045\) so that the system is non-integrable~\cite{kim_testing_2014}.
        As a target state, we aim to prepare a thermal pure quantum state corresponding to the Gibbs ensemble with inverse temperature \(\beta = 0.2\), where the initial state before any quantum control is taken to be a product state \(\Ket{\downarrow\downarrow\cdots\downarrow}\).
        The total preparation time is fixed as \(T=24\) with the time step set as \(\delta t=0.1\), which also determines the total time step to be \(240\).
        
        The set of action candidates \(\{a_t\}\) available for the RL agent is given as
        %of the deep RL at each time step \(t\) is to choose an appropriate unitary from the set %\(\{U=e^{-iG\delta t}\}\) 
        \(\{e^{-iG_k \delta t}\}_{k=1}^6\),
        where the time evolution generator \(G_k\) is chosen from the following six operators:
        \begin{align}
        \begin{split}
            &\ham_{\mathrm{Ising}},\ 
            \sum_{l=1}^LJ\s_{l}^{z}\s_{l+1}^{z}+h\s_{l}^{z},\ 
            g\sum_{l=1}^L\s_{l}^{x},\ 
            \sum_{l=1}^L\s_{l}^{y},\\
            &\sum_{l=1}^L\s_{l}^{x}\s_{l+1}^{y}+\s_{l}^{y}\s_{l+1}^{x},\ 
            \sum_{l=1}^L\s_{l}^{y}\s_{l+1}^{z}+\s_{l}^{z}\s_{l+1}^{y}.
        \end{split}
        \end{align}
        These terms are also used in \ccite{yao_reinforcement_2021}, which discusses how to accelerate the preparation of the ground state using counter-diabatic driving.
        Regarding the local reward \(r_t\), we include the energy density \(L^{-1}\ham_{\mathrm{Ising}}\) and the magnetization density \(L^{-1}\sum_{l}\s_{l}^{z}\).
        Note that these are the sum of local operators acting on two neighboring sites at most.
        
        %Note that the dimension of the arrays \(M\) in \eref{eq:inv_reward} is two.        
        %The set of actions the RL agent choose consists of six unitaries.
        %and we set \(J=1\), \(h=0.9045\), and \(g=0.809\) so that the model is non-integrable.
        %We impose the periodic boundary condition.
        %The target Gibbs ensemble corresponds to inverse temperature \(\beta=0.2\). 
        %The initial state is the product state \(\Ket{\downarrow\downarrow\cdots\downarrow}\).
        %We fix the total time lengths \(T=12\) and \(\delta t=0.1\). The number of time steps is \(120\).
        %We construct the set of unitaries using the generators: \(\ham_{\mathrm{Ising}}\), \(\sum_{i}J\s_{i}^{z}\s_{i+1}^{z}+h\s_{i}^{z}\), 
        %\(g\sum_{i}\s_{i}^{x}\), \(\sum_{i}\s_{i}^{y}\), \(\sum_{i}\s_{i}^{x}\s_{i+1}^{y}\), and \(\sum_{i}\s_{i}^{y}\s_{i+1}^{z}\).
        %The action set that the RL agent can choose is this set of unitaries.

    \subsection{Numerical results}\label{sse:ge_result}
        \begin{figure}[tbp]
            \centering
            \includegraphics[width=\linewidth]{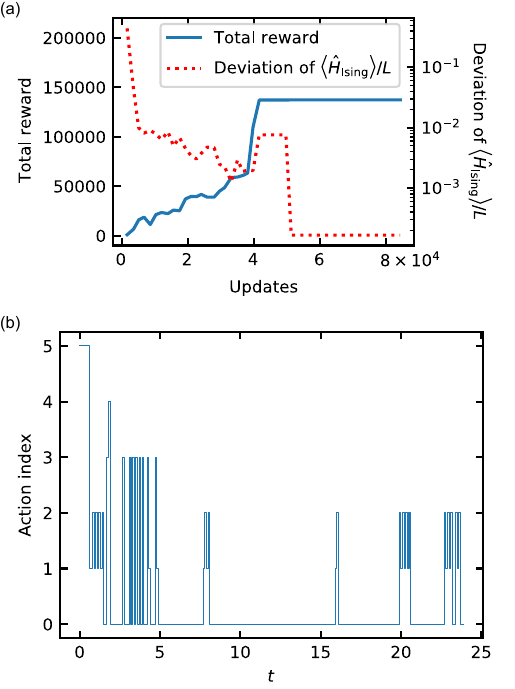}
            \caption{
                (a)~The learning curves for local preparation of the Gibbs ensemble of the transverse field Ising model in the non-integrable regime.
                The blue and red curves denote the total reward and the deviation of the energy density computed for the prepared states, respectively.
                The horizontal axis shows the number of parameter updates.
                %The corresponding model is the transverse field Ising model where the parameter is set so that the model is non-integrable~\eqref{eq:ising}.
                %The deviations were calculated at the end of the episode in the evaluation of the training progress.
                %The lines show the median of the corresponding values in evaluation with random seeds.
                (b)~The optimal action sequence \(\{a_t^*\}\) for local preparation of the Gibbs ensemble  discovered by the RL agent whose training profile is provided in Fig.~\ref{fig:nonint_learning}~(a).
                %The RL-discovered local preparation protocol corresponding to Fig.~\ref{fig:nonint_learning}, in which the Gibbs ensemble is learned.
                %This protocol corresponds to the data in \fref{fig:nonint_learning}.
                The six-fold action indices correspond to
                0:~\(\ham_{\mathrm{Ising}}\),
                1:~\(\sum_{l}J\s_{l}^{z}\s_{l+1}^{z}+h\s_{l}^{z}\),
                2:~\(g\sum_{l}\s_{l}^{x}\),
                3:~\(\sum_{l}\s_{l}^{y}\),
                4:~\(\sum_{l}\s_{l}^{x}\s_{l+1}^{y}+\s_{l}^{y}\s_{l+1}^{x}\),
                5:~\(\sum_{l}\s_{l}^{y}\s_{l+1}^{z}+\s_{l}^{z}\s_{l+1}^{y}\), respectively.
                After quickly manipulating the macroscopic observables to encode the properties of the thermal ensemble, the agent mainly chooses to evolve the system under the target Hamiltonian \(\hat{H}_{\rm Ising}\).
                For all plots, the system size is \(L=16\) with the inverse temperature set as \(\beta = 0.2\).
            }\label{fig:nonint_learning}
        \end{figure}
        
        We now proceed to present numerical results that successfully prepare thermal pure quantum states, which encode both the local observables used during the training and also more non-local ones that are excluded from the reward function.
        Figure~\ref{fig:nonint_learning}~(a) shows an example of the learning curve of the RL agent for \(L=16\).
        We can see that, as the number of training episodes increases, the RL agent learns the better protocols that achieve higher total reward and smaller energy deviation. % [See Appendix.~\ref{app:discovered_protocol} for the concrete protocol.]

        % \section{Control protocol discovered by the RL agent}\label{app:discovered_protocol}

        Here, we briefly discuss the control protocol found by the RL agent.
        Figure~\ref{fig:nonint_learning}~(b) shows the control protocol whose learning curve is shown in Fig.~\ref{fig:nonint_learning}~(a) to realize a thermal pure state described by the Gibbs ensemble. 
        It can be seen that the optimal protocol consists of two stages: first is the sequence of nontrivial actions that change macroscopic observables, and the second is the relaxation of the system under free evolution using \(\hat{H}_{\rm Ising}\), which leads the system to the typical state of \(\hat{H}_{\rm Ising}\). 
        While there is a small number of actions other than the free relaxation at the late stage of the control protocol, we argue that they do not contribute significantly to the result even if we substitute them with \(\hat{H}_{\rm Ising}\) because they consist almost entirely of the alternating use of actions 1 and 2. This can be considered to be effectively equivalent to \(\hat{H}_{\rm Ising}\) in the sense of the Trotter decomposition.

        \begin{figure}[tb]
            \centering
            \includegraphics[width=3.4in]{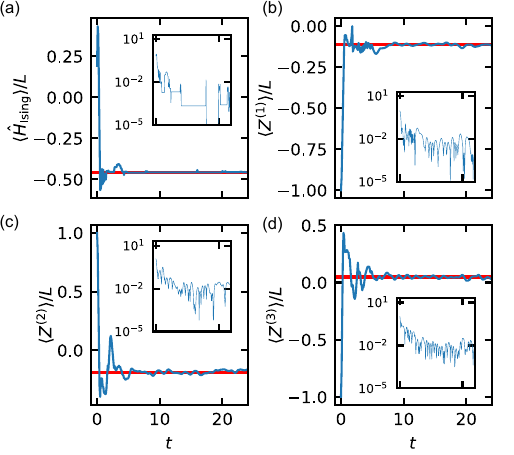}
            \caption{
                (a)--(d)~The dynamics of the local observables generated by the state preparation protocol learned by the deep RL agent, with the insets showing the absolute error from the corresponding expectation values of the target Gibbs state (red line).
                %The red horizontal lines show the corresponding expectation values of the target Gibbs ensemble.
                Each panel displays (a)~energy density \(\ham_{\mathrm{Ising}}/L\),
                (b)~\(Z^{(1)}/L=L^{-1}\sum_{l}\s_{l}^{z}\),
                (c)~\(Z^{(2)}/L=L^{-1}\sum_{l}\s_{l}^{z}\s_{l+1}^{z}\), and
                (d)~\(Z^{(3)}/L=L^{-1}\sum_{l}\s_{l}^{z}\s_{l+1}^{z}\s_{l+2}^{z}\).
                Note that the local observables in (a) and (b) are both included in the reward for the deep RL agent.
                %The insets show the absolute deviations in the expectation values between the states in the protocol and the target Gibbs ensemble.
                %All figures tell that all expectation values converge to the corresponding values of the Gibbs ensemble.
                % We observe that, not only the observables which are included in the RL reward converge to the target value as in (a, b), but also local observables absent in the reward readily converge to the target values as in (c, d).
                We observe convergence not only for physical quantities that are included in the RL reward as in (a) and (b), but also for local observables that are absent from the reward, as in (c) and (d).
                For all plots, the system size is \(L=16\), with the inverse temperature set as \(\beta = 0.2\).
                }
            \label{fig:nonint_dynamics}
        \end{figure}
        %, and its performance converges 
        %as the learning process progresses.

        Figures~\ref{fig:nonint_dynamics}~(a)--(d) display the dynamics of the local observables obtained by the preparation protocol learned by the RL agent.
        Figures~\ref{fig:nonint_dynamics}~(a) and (b) show the observables \(\ham_{\mathrm{Ising}}\) and \(\sum_{l}\s_{l}^{z}\), respectively, which are used for the reward function.
        Both of them converge to the corresponding values of the Gibbs ensemble represented by the red horizontal lines.
        It is more intriguing to see the convergent behaviors in
        Figs.~\ref{fig:nonint_dynamics}~(c) and (d), which show the dynamics of local observables that are not included in the reward function, namely, the two-point correlator \(\sum_{l}\s_{l}^{z}\s_{l+1}^{z}\) and the three-point correlator \(\sum_{l}\s_{l}^{z}\s_{l+1}^{z}\s_{l+2}^{z}\), respectively.
        %Their support is wider than the observables used for the reward. 
        %Nevertheless, both expectation values converge to those of the Gibbs ensemble.
        \begin{figure}[tbp]
            \centering
            \includegraphics[width=\linewidth]{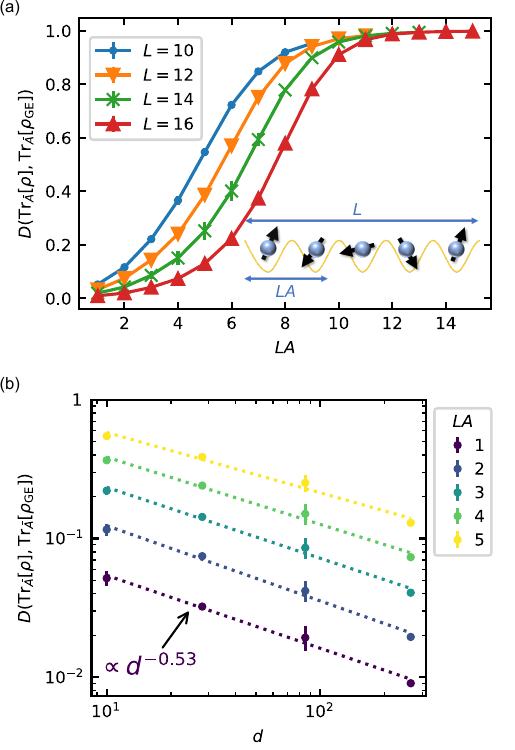}
            \caption{
                %Finite-size scaling with the system size \(L=10, 12, 14, 16\) for the distance function \(D\) between the subsystems of the final states and those of the Gibbs ensemble.
                % Distance of reduced density operator between the target Gibbs state and the prepared state, 
                The distances between the reduced density operators of the target Gibbs state and the prepared state, 
                which are averaged over time and random training instances.
                (a) The error in the distance function \(\bar{D}\)
                at various subsystem sizes \(LA\).
                The blue circles, orange inverted triangles, green crosses, and red normal triangles denote the data for \(L=10, 12, 14, 16\), respectively.
                (b) %The double logarithmic plot to see the finite-size scaling.
                The decay of the error in the reduced density operator along with the corresponding energy shell dimension \(d\), i.e., the number of eigenstates in the energy shell.
                %The horizontal axis shows the Hilbert-space dimension of the system. 
                %The circles represent the result of the RL.
                The dotted lines are guides to the eye showing the fits with \(D=a d^{-b}\),
                where the powers \(b\)  are summarized in \tref{tab:nonint_scaling} of Appendix.~\ref{app:scaling_exp_table}.
                %The power-law decay of the distance functions at small subsystem sizes is compatible with \eref{eq:Typicality_scaling} and the scaling of the variance of observables \(\sigma_{\mathrm{ETH}}\) in non-integrable systems (\(\propto d^{-0.5}\))~\cite{beugeling_finite-size_2014}.
                %the variance of observables \(\sigma_{\rm ETH}\) under strong ETH in non-integrable systems~\cite{beugeling_finite-size_2014}. %as a power law with the Hilbert-space dimension as far as the subsystem sizes are small.
                The error bars correspond to the standard deviations of different protocols learned independently with random seeds.
                In both plots, the system is controlled by the deep RL agent at \(0\leq t\leq 24\), and then undergoes free evolution \(\exp [ -i\ham_{\mathrm{Ising}} t ]\) in \(24<t\leq 48\). 
                Note that the distance \(\bar{D}\) concerns the average over the free evolution period and also the random training instance. %seeds for \(L=10, 12, 14, 16\), respectively.
                }\label{fig:nonint_scaling}
        \end{figure}

        It is natural to wonder how accurately the general observables, or the reduced density operators of the local systems, are captured by the local preparation protocol.
        For this purpose, we evaluate the distance between the reduced density operators of the prepared pure state and the target Gibbs ensemble as
        \begin{align}
            D(\rho,\rho')\coloneqq\frac{\left\lVert \rho-\rho'\right\rVert _{\mathrm{F}}}{\sqrt{\left\lVert \rho\right\rVert_{\mathrm{F}}^{2}-\left\lVert \rho'\right\rVert_{\mathrm{F}}^{2}}},\label{eq:distance}
        \end{align}
        where \(\left\lVert A\right\rVert _{\mathrm{F}} \coloneqq\sqrt{\tr\left\lbrack A^{\dagger}A \right\rbrack }\) is the Frobenius norm~\footnote{Note that the chosen distance function can be efficiently computed for Slater determinants, which is used to express free-fermionic states obtained from Jordan-Wigner transformation. 
        See \ccite{fagotti_reduced_2013} for the detailed property of this distance. }. 
        
        We can see from Fig.~\ref{fig:nonint_scaling}~(a) that the time average of the distance function given in \eref{eq:distance}, which we denote as \(\bar{D}\), is suppressed for the smaller subsystem size \(LA\). What is more interesting is the scaling with respect to the total system size; we observe nontrivial suppression of error along with the increase of the system size \(L\).

        We go further into the scaling of the error suppression by performing finite-size scaling on the averaged distance function as shown in Fig.~\ref{fig:nonint_scaling}~(b). 
        %is the the scaling law with the corresponding energy shell dimension \(d\).
         %in \fref{fig:nonint_scaling} (b).
        Here, we observe that the scaling of the suppression is given as 
        \begin{equation}
         \bar{D}=\calO(d^{-b}),\label{eq:gibbs_scaling}   
        \end{equation}
        where \(d\) is the dimension of the corresponding energy shell. 
        To be specific, \(d\) is obtained by counting the number of eigenstates included within the energy window \([ \braket{\ham_{\mathrm{Ising}}}_{\beta}-\varepsilon_{\mathrm{e}}L,~\braket{\ham_{\mathrm{Ising}}}_{\beta} ]\), where \(\braket{\ham_{\mathrm{Ising}}}_{\beta}\) is the energy expectation value of the target Gibbs ensemble at inverse temperature \(\beta\),
        %~\footnote{Note that \(\|\ham_{\mathrm{Ising}}\|=O(L)\)}
        and the shell width is fixed as \(\varepsilon_{\mathrm{e}}=0.5\) (for further discussion on the choice of \(\varepsilon_{\mathrm{e}}\), see \aref{app:shell_dep}).

       The scaling of the distance given in \eref{eq:gibbs_scaling} implies that \(\bar{D}\) decays exponentially with the system size \(L\).
       We argue that this is compatible with the scaling of canonical typicality, given in \eref{eq:Typicality_scaling}. 
       %Namely, we may attribute such a scalability of the preparation of the protocol to the typicality of eigenstates in non-integrable systems, in which the fluctuation of local observables are suppressed exponentially as \(\sigma_{\mathrm{ETH}} = \calO( d^{-0.5})\)~\cite{beugeling_finite-size_2014}.
       This feature supports an expectation that our local preparation protocol for the Gibbs ensembles becomes exponentially more precise as the system size increases.

       As a technical remark, we mention that the dimension shown in the figure corresponds to the size of the symmetry-resolved Hilbert space, namely the parity and the momentum.

\section{Application to generalized Gibbs ensemble}\label{sss:gge}
    \subsection{Model and setup}
        %The second target is GGEs\@.
        We next describe an even more intriguing system that is integrable so that the thermodynamic ensembles is described by GGEs\@.
        Specifically, we consider the XX model on a periodic chain with a longitudinal field:
        \begin{align}
            \ham_{\mathrm{XX}}=-\sum_{l=1}^{L}\left\lbrack  \frac{J}{2}\left( \s_l^x\s_{l+1}^x+\s_l^y\s_{l+1}^y \right)+h\s_{l}^{z} \right\rbrack ,\label{eq:XX}
        \end{align}
        where \(J\) is the amplitude of interaction, \(h\) is the strength of the magnetic field along the \(z\)-axis and \(L\) is the system size.
        In the following, the parameters are fixed as \(J=1\) and \(h=2\). 
        
        %We impose the periodic boundary condition.
        
        The integrability of the XX model can be verified straightforwardly by mapping into a non-interacting fermionic system via the Jordan-Wigner transformation:
        \begin{eqnarray}
            \ham_{\mathrm{XX}} &=&-\sum_{l=1}^{L}\lbrack J( \ha_{l}^{\dagger} \ha_{l+1}+\ha_{l+1}^{\dagger} \ha_{l} )
            +2h( \ha_{l}^{\dagger} \ha_{l} -\frac{1}{2}) \rbrack ,\label{eq:tight-bind}\\
            &=&-\sum_{k}\left\lbrack 2\left\lparen  J\cos k +h \right\rparen\hn_{k}-h \right\rbrack ,\label{eq:tight-bind_diag}       
        \end{eqnarray}
        where \(\ha_{l}^{\dagger}\) and \(\ha_{l}\) are fermionic creation and annihilation operators 
        that are related to the Pauli operators as \(\hat{a}_l^{(\dag)} = \prod_{l'<l}\hat{\sigma}_{l'}^z \hat{\sigma}_l^{+(-)}\), where \(\hat{\sigma}_l^{\pm} = (\hat{\sigma}_l^x \pm i\hat{\sigma}_l^y)/2\).
        In the second row of \eref{eq:tight-bind_diag}, we move into the Fourier space by introducing the mode occupation operator \(\hn_{k}=\ha_{k}^{\dagger} \ha_{k}\) for \(\ha_{k}=L^{-1/2}\sum_{l}\ha_{l}e^{-ikl}\).
        %This is block diagonal in the Fourier space as
        %\begin{eqnarray}
            %\ham_{\mathrm{tb}} &=&-\sum_{k}\left\lbrack 2\left\lparen  J\cos k +h \right\rparen\hn_{k}-h \right\rbrack ,\label{eq:tight-bind_diag},
        %\end{eqnarray}        
        %where \(\hn_{k}=\ha_{k}^{\dagger} \ha_{k}\) is the mode occupation operator for \(\ha_{k}=L^{-1/2}\sum_{l}\ha_{l}e^{-ikl}\).
        We can construct as many LIOMs as the system size \(L\) by taking the linear combination as
        \begin{align}
            \hat{I}_{n}^{+}&=-2J\sum_{k}\cos{(nk)}\hat{a}_{k}^{\dagger} \hat{a}_{k}
            =-J\sum_{l}\left\lparen  \hat{a}_{l}^{\dagger} \hat{a}_{l+n}+\hat{a}_{l+n}^{\dagger} \hat{a}_{l} \right\rparen, \nonumber\\
            \hat{I}_{n}^{-}&=-2J\sum_{k}\sin{\left\lparen  nk \right\rparen}\hat{a}_{k}^{\dagger} \hat{a}_{k}
            =iJ\sum_{l}\left\lparen  \hat{a}_{l}^{\dagger} \hat{a}_{l+n}-\hat{a}_{l+n}^{\dagger} \hat{a}_{l} \right\rparen.\nonumber
        \end{align}
        For convenience in the later discussion, we mention the rightmost sides to remark that  \(\hat{I}_n^{\pm}\) can be explicitly expressed as sums over hopping terms between the \(n\)-th nearest neighboring sites in the fermionic picture.

        As the target thermodynamic ensemble, we aim for the GGE constructed from the LIOMs as 
        %We set the target GGE\@:
        \begin{align}
            \rho_{\mathrm{GGE}}=\frac{\exp\left\lbrack -\sum_{n}\sum_{\sigma=\pm}\lambda_{n}^{\sigma}\hat{I}_{n}^{\sigma} \right\rbrack }{\tr \left\lbrack \exp\left\lbrack -\sum_{n}\sum_{\sigma=\pm}\lambda_{n}^{\sigma}\hat{I}_{n}^{\sigma} \right\rbrack  \right\rbrack },
        \end{align}
        where \(\{\lambda_{n}^{\sigma}\}\) are the Lagrange multipliers. 
        More precisely, we set the initial state of the control system to be the ground state of \(\ham_{\mathrm{tb}}\) in the space of the total particle number \(N_{\mathrm{f}}\), and aim to prepare a prethermal pure state corresponding to a given set of Lagrange multipliers \(\{\lambda_{n}^{\sigma}\}\)~(for details regarding the parameter choice of \(\{\lambda_{n}^{\sigma}\}\), see \aref{app:lagrange}).
        The total preparation time is fixed as \(T=40\) and the time step as \(\delta t=0.2\); thus the total number of time steps is \(200\).

        %The initial state is the ground state of \(\ham_{\mathrm{tb}}\) in the space of the total particle number \(N_{\mathrm{f}}\).
        %, i.e., all modes under the Fermi energy are occupied.
        
        In parallel with the case for Gibbs ensembles,
        we allow the deep RL agent to choose \(a_t\) as an appropriate unitary from a set \(\{e^{-iG_k\delta t}\}_{k=1}^{15}\), where the time evolution generator \(G_k\) is chosen from the following 15 operators:
        %At each time step, the agent of the deep RL chooses an appropriate unitary from the set \(\{U=e^{-iG\delta t}\}\) where the time evolution generator \(G\) is chosen from the following:
        \begin{align}
            \begin{split}
                &\ham_{\mathrm{XX}},~
                \sum_{l=1}^{L}\ha_{l}^{\dagger} \ha_{l}\cos \frac{m\pi}{L}l,~
                \sum_{l=1}^{L}\ha_{l}^{\dagger} \ha_{l}\sin \frac{m\pi}{L}l,\\
                &\sum_{l=1}^{L}\left\lparen  \ha_{l}^{\dagger} \ha_{l+j}+\ha_{l+j}^{\dagger} \ha_{l} \right\rparen\cos \frac{n\pi}{L}l,\\
                &\sum_{l=1}^{L}\left\lparen  \ha_{l}^{\dagger} \ha_{l+j}+\ha_{l+j}^{\dagger} \ha_{l} \right\rparen\sin \frac{n\pi}{L}l,
            \end{split}
        \end{align}
        where \(m\in \{2, 4, L\}\), and \(\left\lparen  j, n \right\rparen \in \{\left\lparen 1, 2 \right\rparen, \left\lparen  2, 2 \right\rparen, \left\lparen 1, L \right\rparen, \left\lparen  2, L \right\rparen \}\).
        Note that these operators are chosen so that they are not diagonal in the position or momentum basis.

        Regarding the local reward \(r_t\), we employ normalized LIOMs \({\lbrace \hat{I}_{n}^{+}/L\rbrace}_{1 \leq n \leq n_{\mathrm{local}}}\), where we fix \(n_{\rm local} = 4\) in the following.
        All \(\hat{I}_n^-\) terms are excluded since they are constantly zero not only for the initial and target states, but also for  any intermediate states \(\rho_t\) evolved with the above unitaries.
        
        %Therefore, the set of actions the RL agent choose consists of fifteen unitaries.

        %We consider using the LIOM for the reward function.
        %\(\hat{I}_{0}^{+}\) is proportional to the total particle number and constant in this work.
        %In addition, we verified \(\braket{\hat{I}_{n}^{-}}=0\) for all initial states, preparing states using the protocol found by RL and target states.
        %Thus, we use only \({\lbrace L^{-1}\hat{I}_{n}^{+}\rbrace}_{n=1,\ldots, n_{\mathrm{local}}}\) for the reward function.
        %In the following, we fix \(n_{\mathrm{local}}=4\), and the dimension of the arrays \(M\) in \eref{eq:inv_reward} is four.
    \subsection{Numerical results}\label{sse:gge_result}
        \begin{figure}[tbp]
            \centering
            \includegraphics[width=\linewidth]{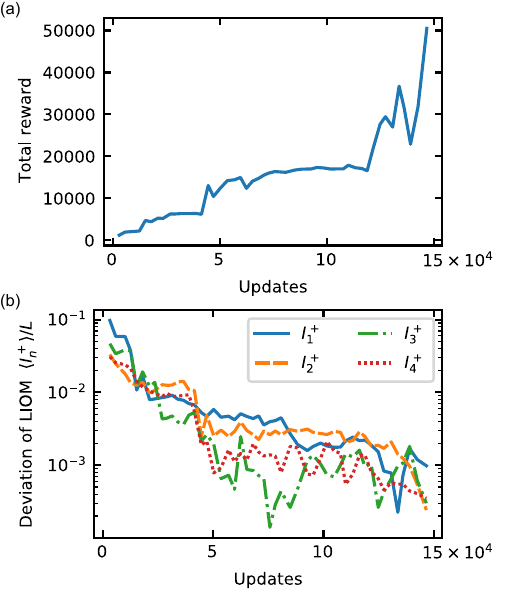}
            \caption{   
                The learning curves for local preparation of the GGE\@.
                These figures show the total rewards and the deviations of the LIOM densities at the end of the episodes in evaluating the training progress.
                The horizontal axis shows the number of parameter updates.
                (a) The total rewards. (b) The deviations of the LIOM densities.
                The corresponding model is the transverse field Ising model where the parameter is set so that the model is integrable.
                We add the LIOMs \(\hat{I}_{n}^{+}\) with separation \(n=1, 2, 3, 4\) to the reward of RL\@.
                The system size \(L=120\). 
                %The target expectation values of LIOMs correspond to an eigenstate chosen randomly with the Fermi distribution where the inverse temperature \(\beta = 0.4\)
                %and the chemical potential are obtained by optimizing so that the number of fermions \(N_{\mathrm{f}}=10\).
                %Only eigenstates with \(N_{\mathrm{f}}=10\) are accepted.
                %The deviations were calculated at the end of the episode in evaluation of the training progress.
                The lines show the median of the corresponding values in evaluation with random seeds.
            }\label{fig:int_learning}
        \end{figure}

        We now present the numerical results obtained by running the deep RL algorithm to prepare prethermal pure quantum states that capture the characteristics of the GGE\@.
        As we show in the learning curve in \fref{fig:int_learning}(a), the deep RL agent successfully learns to improve the local preparation protocol. 
        This can be more quantitatively understood from  Fig.~\ref{fig:int_learning}(b), which evaluates the absolute difference in the expectation values of LIOMs between the target GGE and the prepared state.
        %The learning curve of the deep RL protocol for the GGE is shown in 
        %The agent learns the protocol for the system size \(L=120\), \(N_{\mathrm{f}}=10\), and \(n_{\mathrm{local}}=4\).
        %\fsref{fig:int_learning} show (a) the total rewards and (b) the deviations of the LIOMs at the end of the episodes in evaluating the training progress
        %and tell that the agent finds the protocol with a high total reward and slight deviations of LIOMs.
        \begin{figure}[tbp]
            \centering
            \includegraphics[width=\linewidth]{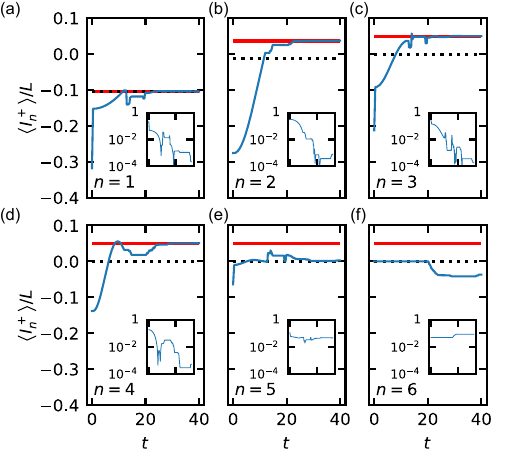}
            \caption{
                The expectation values of the LIOMs in the protocol found by RL for the GGE (blue).
                Each panel shows the LIOM with different separations.
                The red horizontal lines show the corresponding expectation values of the target GGE\@.
                The dotted black lines represent the relevant Gibbs ensemble, which shares the same expectation values of the total particle number and the energy.
                The insets show the absolute deviations in the expectation values \(\langle \hat{I}_{n}^{+}\rangle/L\) between the states in the protocol and the target GGE\@.
                We add the LIOMs \(\hat{I}_{n}^{+}\) with separation \(n=1, 2, 3, 4\) to the reward of RL\@.
                Only LIOMs with separation \(n\leq 4\) converge to the expectation values of the GGE\@.
            }\label{fig:int_dynamics_LIOM}
        \end{figure}

        Let us further investigate the dynamics of the LIOMs generated by the preparation protocol discovered by the RL agent\@. 
        As shown in Fig.~\ref{fig:int_dynamics_LIOM},
        %Each figure shows the LIOM with a different separation \(n\).
        the behavior of the LIOMs seem to be qualitatively different depending on \(n\) in the sense that, the only LIOMs with separation \(n\leq n_{\mathrm{local}}=4\), which we add to the reward, seem to converge to the expectation values of the GGE\@.
        
        \begin{figure}[tbp]
            \centering
            \includegraphics[width=\linewidth]{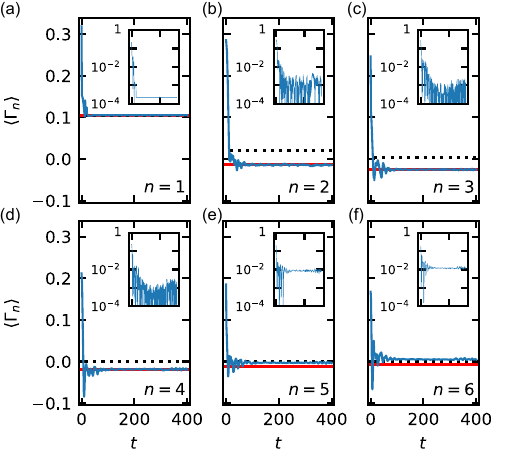}
            \caption{
                The relaxation processes of the expectation values of the correlation functions given in \eref{eq:correlation}, which are non-conserved quantities of \(\ham_{\mathrm{XX}}\)~\eqref{eq:XX} (blue).
                %The blue lines represent the expectation values the quantum state in the protocols.
                Each panel shows the correlation function with different separation \(n\), which means the correlation function corresponds to the two spins separated by \(n-1\) sites.
                The red horizontal lines show the corresponding expectation values of the target GGE\@,
                whereas the dotted black lines represent the relevant Gibbs ensemble that shares the same expectation values of the total particle number and the energy.
                Note that the system is controlled with the learned protocol in \(0\leq t \leq 40\), and subsequently undergoes free evolution with \(\ham_{\mathrm{XX}}\).
                The only correlation functions with separation \(n\leq 4\) fluctuate around the expectation values of the GGE\@.
                Note that the correlation function with \(n=1\), where the Jordan-Wigner string does not appear, is proportional to the LIOM with \(n=1\), i.e., it is the conserved quantity of \(\ham_{\mathrm{XX}}\).
            }\label{fig:int_dynamics_correlation}
        \end{figure}
        Similar behavior can also be observed for non-conserved quantities such as the correlation function
        \begin{align}
            \Gamma_{n}\coloneqq\frac{1}{L}\sum_{l}\left\lparen  \s_{l}^{+}\s_{l+n}^{-}+\s_{l+n}^{+}\s_{l}^{-} \right\rparen,\label{eq:correlation}
        \end{align}        
        which can be seen as a representative of the local operators acting on neighboring \(n+1\) sites.
        As we can see in Fig.~\ref{fig:int_dynamics_correlation}, there is a notable convergence into the GGE values for \(n \leq n_{\rm local}\), while the errors seem to remain for \(n > n_{\rm local}\).
        
        \begin{figure}[tbp]
            \centering
            \includegraphics[width=\linewidth]{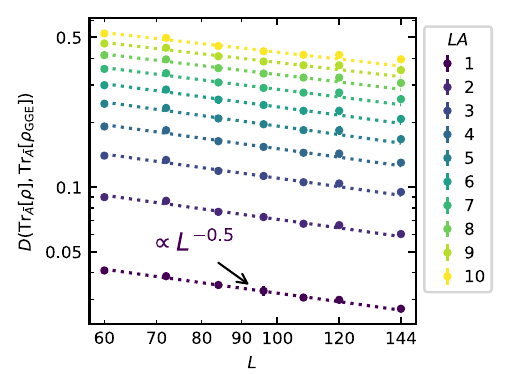}
            \caption{
                Finite-size scaling of the distance function of the reduced density operators between the prepared states and the target GGE\@.
                %The horizontal axis shows the system size.
                %The circles represent the result of the RL\@.
                The dotted lines are guides to the eye that show the power-law fitting \(\bar{D} = \calO(L^{-b})\), where the powers \(b\) are summarized in Table~\ref{tab:int_scaling} of Appendix.~\ref{app:scaling_exp_table}.
                %the fits with \(D=a L^{b}\) for the system size \(L = 60, 72, 84, 96, 108, 120\).
                %The distance functions decay as a power law with the system size.
                The error bars correspond to the standard deviations of different protocols learned independently with random seeds.
                The system is controlled by the deep RL agent at \(0\leq t \leq 40\), and then undergoes free evolution by \(\ham_{\mathrm{XX}}\). Note that the distance \(\bar{D}\) concerns the average over time, i.e., the free evolution period, and the random training instance. The number of random seeds is five at most.
                %This result cannot rule out the case that the distance between the target GGE and relevant Gibbs ensemble decays, and we just prepare the Gibbs ensemble because I choose an eigenstate randomly when I decide the target LIOM expectation values. 
                %In the context of the ETH, with only the energy, almost all eigenstates coincide with the Gibbs ensemble, and a power law may emerge in the scaling of the variance of the expectation values.
                %Now, I perform the additional analysis and calculation.
                }\label{fig:int_scaling}
        \end{figure}

        Furthermore, we focus on the scaling of the error suppression. %We argue that the typicality again plays an important role.
        As we can observe from the finite-size scaling of the averaged distance function \(\bar{D}(\rho_t, \rho_{\rm GGE})\) in Fig.~\ref{fig:int_scaling}, the distance between the reduced density operators is suppressed as \(\bar{D} = \mathcal{O}(L^{-b})\).
        This behavior is compatible with the scaling of the fluctuation of local observables in the GME, given in \eref{eq:gme_decay}.

        Here, we conjecture that while the errors for \(LA\lesssim n_{\rm local} +1\) are suppressed polynomially even for larger system sizes, the errors for \(LA \gtrsim n_{\rm local}+1\) may saturate at finite values. 
        This is because the current local preparation scheme learns the LIOMs \(\hat{I}_n^\sigma\) with \(n \leq n_{\rm local}\) to encode the prepared state into the ``LIOM shell" only for such conserved quantities.
        This means that the prepared state fully encodes the macroscopic properties of the tGGE but not those of the GGE\@.
        We numerically find that the distance between the tGGE and the GGE remains finite even if the total system is at the thermodynamic limit~[See Appendix~\ref{app:tgge_gge_distance}], which is in agreement with Ref.~\cite{fagotti_reduced_2013}, which has investigated the integrable region of the  transverse-field Ising chain.
        This supports our conjecture that the distance between the GGE and the local-prepared state is not suppressed in the thermodynamic limit.
        Meanwhile, it is possible that the error from the tGGE itself is suppressed polynomially.

\section{Conclusion}\label{sss:conclusion}

    In this work, we propose a deep-RL-based quantum state preparation framework for thermodynamic ensembles that relies solely on a few local observables but not on global features such as the fidelity. 
    The core idea is to leverage the typicality of pure states in quantum many-body systems; the macroscopic properties can be encoded simply via learning a few local observables and undergoing free evolution.
    We provid numerical demonstrations in which the deep RL agent is successfully trained to learn the macroscopic properties of the Gibbs ensembles~(\fref{fig:nonint_learning}) and the GGEs~(\fref{fig:int_learning}). 
    We find that the accuracy of the prepared state improves exponentially with the system size for the former~(\fref{fig:nonint_scaling}) and polynomially for the latter~(\fref{fig:int_scaling}), which is consistent with the argument of typicality within a given shell of local conserved quantities.
    
    We envision four future directions for our work.
    First, the application to interacting integrable models is an important issue for local preparation.
    This issue is related to the previous work on which conserved quantities should be considered to predict the local properties of the steady state in interacting integrable systems, which can be solved using the Bethe ansatz~(see, e.g., \ccite{ilievski_complete_2015}).
    Based on the prior work, we can expect to be able to perform local preparation if we also include \emph{quasi-local} conserved quantities in the reward.
    % However, to consider using typicality in integrable systems, for the system sizes where exact diagonalization is possible, the finite size effect is significant, i.e., 
    % even if the local conserved quantities are made close to the target ensemble values, the fluctuations in the expected values of local observables become large.
    % To overcome this problem, we simulate non-interacting integrable systems at the larger size by exploiting the fact that the XX model can be mapped to a free fermionic system.
    % In order to take advantage of typicality in interacting integrable systems, new and sophisticated simulation methods, which enable us to simulate large interacting integrable systems efficiently, may be needed. 
    Note that in such systems, the finite-size effect on the fluctuation of local observables is severe in system sizes that are tractable by exact diagonalization.
    How to simulate interacting integrable systems efficiently in a scalable way so that a local preparation strategy can be pursued is an open problem.

    The second important question is the generalization of the local preparation protocol to include, e.g., dissipative terms, measurement and feedback, or postselection. We naturally expect that the powerful explorability of the deep RL framework is not limited to coherent control but could be applied to broader operation sets. 
    
    Third, we may consider the application of the local preparation protocol for the task of Hamiltonian learning~\cite{rudinger_compressed_2015,bairey_learning_2019,anshu_sample-efficient_2021} by attempting to encode the macroscopic properties using unitaries that do not explicitly contain information about the Hamiltonian itself.
    
    Finally, it is intriguing to investigate how the local preparation protocol is affected by various noises, such as the statistical noise that accompanies sampling over observables.
    Efficient estimation methods such as randomized measurement schemes~\cite{huang_predicting_2020} will be essential to boost the training accuracy of the RL agent.

\section{Acknowledgments}
    The authors wish to thank fruitful discussion with Ryusuke Hamazaki and Jiahao Yao.
    S. B. is supported by Materials Education rogram for the future leaders in Research, Industry, and Technology (MERIT) of The University of Tokyo.
    N.Y. wishes to
    acknowledge JST PRESTO No. JPMJPR2119 and JST Grant Number JPMJPF2221.
    Y.A. acknowledges support from the Japan Society for the Promotion of Science (JSPS) through Grant Nos. JP19K23424 and JP21K13859.
    T.S. is supported by JSPS KAKENHI Grant Number JP19H05796,
    JST CREST Grant Number JPMJCR20C1, Japan,
    and JST ERATO-FS Grant Number JPMJER2204, Japan.
    N.Y. and T.S. are also supported by Institute of AI and Beyond of
    The University of Tokyo.
    We implemented the exact time evolution in \sref{sss:ge} with QuSpin~\cite{weinberg_quspin_2017,weinberg_quspin_2019} and 
    the RL algorithm with PyTorch~\cite{NEURIPS2019_9015}.
    The RL in \sref{sss:gge} is performed on AI Bridging Cloud Infrastructure (ABCI) of National Institute of Advanced Industrial Science and Technology (AIST).

\appendix
\renewcommand\thefigure{\thesection\arabic{figure}}    
\setcounter{figure}{0}

\section{Layout of the neural network and hyperparameters}\label{app:layout}

    \begin{figure}[tb]
        \centering
        \includegraphics[width=\linewidth]{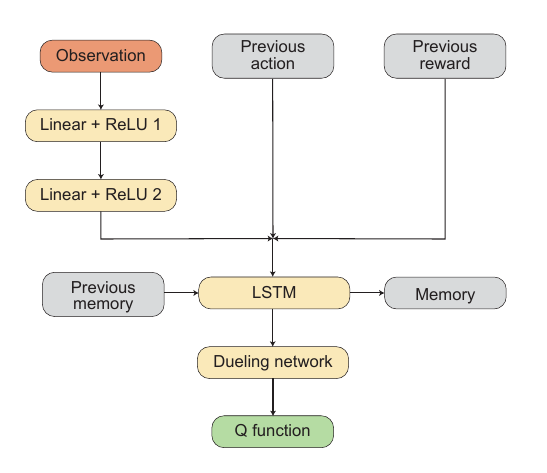}
        \caption{
            The abstract deep NN architecture used for RL in the present work.
            At each time step \(t\), the deep NN takes as input the observation \(o_t\), the previous action \(a_{t-1}\), and the previous reward \(r_{t-1}.\)
            After intermediate computations by fully connected layers, LSTM, and the dueling network, the deep NN outputs the estimate of the Q function.
            %The input is the observation, the previous action, and the previous reward.
            %In this work, we choose the action history \(o_{t}=\left\lparen  a_{0}, a_{1},\ldots,a_{t-1},-1,\ldots,-1 \right\rparen\) as the observation at time step \(t\).
            %The output is the action-value function (Q function) corresponding to \eref{eq:Q}.
            %Between the input layer and output layer, there are three hidden layers and a network: two layers with the activation function (rectified linear unit, ReLU), an LSTM layer, and a dueling network~\cite{wang_dueling_2016}.
            %The observation is fed to two hidden layers with the activation function (rectified linear unit, ReLU).
            %The output of these hidden layers is concatenated
            }\label{fig:RL_architecture}
    \end{figure}

    \begin{table}[tb]
        \centering
        \begin{tabular}{l|r|r}\toprule
            &\sref{sss:ge}&\sref{sss:gge}\\\midrule
            %Input & (120,6,1)& (200,15,1)\\
            Linear + ReLU 1 & \(240\rightarrow512\)& \(200\rightarrow1024\)\\
            Linear + ReLU 2 & \(512\rightarrow512\)& \(1024\rightarrow1024\)\\
            LSTM & \((512,6,1)\rightarrow512\)& \((1024,15,1)\rightarrow1024\)\\
            Dueling network & \(512\rightarrow6\)& \(1024\rightarrow15\)\\
            \bottomrule
        \end{tabular}
        \caption{
            The input and output sizes of each layer are shown as the values preceding and following the arrows, respectively. 
            The input of the NN consists of the action history, the one-hot representation of the previous action, and the previous reward.
            The input of the ``Linear + ReLU 1'' layer is only the action history. The one-hot representation of the previous action and the previous reward is concatenated with the output of ``Linear + ReLU 2'' before the LSTM layer.
            }\label{tab:NN_layer}
    \end{table} 
    
    \begin{table}[tb]
        \centering
        \begin{tabular}{ll|r}\toprule
            %&\\\midrule
            Reward discount \(\gamma\)&&0.997\\
            Minibatch size && 324(\sref{sss:ge})\\
            &&380(\sref{sss:gge})\\
            Sequence length&&40\\
            %burn-in length&&40\\
            Optimizer&& Adam~\cite{kingma_adam_2015}\\
            Optimizer setting &Learning rate&\(10^{-4}\)\\ 
            &\(\varepsilon\)&\(10^{-3}\)\\
            &\(\beta\)&\((0.9, 0.999)\)\\
            Replay ratio&&1\\
            %store_rnn_state_interval=40
            Gradient norms clip&&80\\
            %n_step_return=5
            %pri_alpha=0.9
            %pri_beta_init=0.6
            %pri_beta_final=0.6
            %priority exponent&0.9\\
            %input_priority_shift=2
            %value_scale_eps=1e-3,  # 1e-3 (Steven).
            \bottomrule
        \end{tabular}
        \caption{
            The hyperparameters used in the NNs.
            The agent performs updates on batches of \((\text{minibatch size}\times\text{sequence length})\) observations.
            The replay ratio means the effective number of times that each experienced observation is being replayed for training purposes.
            For the details of the hyperparameters, see \ccite{kapturowski_recurrent_2019} and its previous non-LSTM version, \ccite{horgan_distributed_2018}.
            The other parameters follow thosed used in \ccite{stooke_rlpyt_2019}.
            %\ccite{kapturowski_recurrent_2019}.
            }\label{tab:NN_hyper}
    \end{table} 

    %The architecture of the NN is shown in \fref{fig:RL_architecture}.
    %As we show the architecture in \fref{fig:RL_architecture},
    Here, we describe the architecture of the deep NN that is used to estimate the action-value function (Q function). 
    Figure~\ref{fig:RL_architecture} shows the overall picture;
    the input for the LSTM at time step \(t\) is the observation \(o_t\), the previous action \(a_{t-1}\), and the previous reward \(r_{t-1}\), whereas the output is Q function whose optimal expression is given in \eref{eq:Q} in the main text. 
    Refer to \tref{tab:NN_layer} for the size of the input and output of each layer and \tref{tab:NN_hyper} for the hyperparameter used in the NN\@.
    In the following, we further describe the details of the structure.

    The observation \(o_t\) is chosen to be 
    the action history \(o_{t}=\left\lparen  a_{0}, a_{1},\ldots,a_{t-1},-1,\ldots,-1 \right\rparen\), which
    %The 
    %The observation 
    is fed to the fully connected layers.
    fully connected layers first perform linear transformation, and then apply a non-linear activation function which is chosen to be the rectified linear unit (ReLU) in the present work.
    %two hidden layers with the activation function (rectified linear unit, ReLU).
    %These two hidden layers perform matrix multiplications added biases, i.e., \(\alpha AB+\beta C\), where \(A\) is an input tensor, \(B,~C\) are tensors, and \(\alpha\), \(\beta\) are scalar.
    The intermediate output from the second fully connected layer is concatenated with the previous choice of action \(a_{t-1}\) and the reward in the previous time step \(r_{t-1}\), and then fed to the LSTM layer.
    %The output of these hidden layers is concatenated with the previous action (one-hot representation) and reward and then fed to the LSTM layer.

    The LSTM layer is introduced so that the network can refer to the history of the computational results at \(t'<t\) to estimate the Q function at time step \(t\). 
    Namely, the input of the LSTM layer is not only the one mentioned above but also its "memory", including a \emph{hidden state} (short-term memory) and a \emph{cell state} (long-term memory)~\cite{hochreiter_long_1997}.
    Refer to literature such as Ref.~\cite{sherstinsky_fundamentals_2020} for detailed information.
    %depends not only on the input of the following dueling network~\cite{wang_dueling_2016}, but also its memory, including a \emph{hidden state} (short-term memory) and a \emph{cell state} (long-term memory).
    This output memory is fed to the LSTM layer in the next time step \(t\), which enables the deep NN to successfully deal with time-series inputs.
    
    %Therefore, the NN utilizes time-series inputs.

    In the subsequent dueling network~\cite{wang_dueling_2016}, the input is separated into two branches. 
    One branch evaluates the value of the observation \(V(o)\), and the other branch evaluates the advantage of actions regarding the observation \(A(o,a)=Q(o,a)-V(o)\).
    The output Q function of the dueling network is obtained by summing the outputs of the two branches: \(Q(o,a)=V(o)+A(o,a)\). 
    This separation may contribute to better training stability, faster convergence, and better performance.
    %The hyperparameters used in the NN are summarized in .

    As a computational resource, we use a single CPU and four GPUs (Intel Xeon E5-2698 v4, \(4 \times\)NVIDIA Tesla V100) in \sref{sss:ge} and
    two CPUs and four GPUs (\(2 \times\)Intel Xeon Gold 6148, \(4 \times\)NVIDIA Tesla V100) in \sref{sss:gge}, respectively.

    \section{Fitting parameters for the finite-size scaling of the preparation accuracy}\label{app:scaling_exp_table}
    In Tables~\ref{tab:nonint_scaling} and \ref{tab:int_scaling}, we summarize the powers \(b\) obtained by the fit for the finite-size scaling of the local preparation accuracy in \sref{sse:ge_result} and \sref{sse:gge_result}, respectively.
    %\tref{tab:nonint_scaling} shows the powers for the result of the finite-size scaling regarding the Gibbs ensembles in \sref{sse:ge_result}.
    %\tref{tab:int_scaling} shows the powers for the result of the finite-size scaling regarding the GGE in \sref{sse:gge_result}.
    
    \begin{table}[tb]
        \centering
        \begin{tabular}{lr}\toprule
            Subsystem size \(LA\)&Power \(b\)\\\midrule
            1&0.53(4)\\
            2&0.54(4)\\
            3&0.51(4)\\
            4&0.48(4)\\
            5&0.43(4)\\
            %5&0.37(4)\\
            %6&0.31(4)\\
            %7&0.22(4)\\
            %8&0.13(3)\\
            %9&0.06(2)\\
            \bottomrule
        \end{tabular}
        \caption{
            The powers obtained by the fit for the result of the finite-size scaling of the distance function \(\bar{D}\), given by \eref{eq:distance}, between the prepared states and the target Gibbs ensembles for the system size \(L=10, 12, 14, 16\) in \sref{sse:ge_result}.
        }\label{tab:nonint_scaling}
    \end{table}

    \begin{table}[tb]
        \centering
        \begin{tabular}{lr}\toprule
            subsystem size \(LA\)&power \(b\)\\\midrule
            1&0.50(3)\\
            2&0.50(5)\\
            3&0.46(4)\\
            4&0.43(4)\\
            5&0.41(4)\\
            6&0.41(3)\\
            7&0.41(3)\\
            8&0.41(3)\\
            9&0.41(3)\\
            10&0.41(3)\\\bottomrule
        \end{tabular}
        \caption{
            The powers obtained by the fit for the result of the finite-size scaling with the system size \(L = 60, 72, 84, 96, 108, 120\) for the distance function \(\bar{D}\), given by \eref{eq:distance}, between the prepared states and the target GGE in \sref{sse:gge_result}.
            }\label{tab:int_scaling}
    \end{table}

\section{Relationship between energy shell width \(\varepsilon_{\rm e}\) and scaling of preparation accuracy}\label{app:shell_dep}
\begin{figure}[tb]
        \centering
        \includegraphics[width=3.4in]{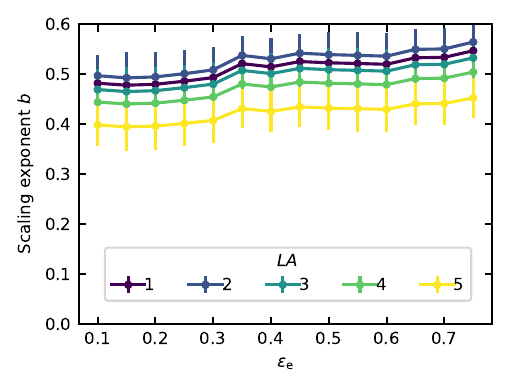}
        \caption{      
        The relationship between the energy shell width \(\varepsilon_{\rm e}\) and the scaling exponent \(b\) of the distance \(\bar{D}\), with regard to the local preparation of the Gibbs ensemble. The parameters of the system are identical to those used in Sec.~\ref{sse:ge_result}.
        }\label{fig:shell_dep}
            %The energy  \(\varepsilon_{\mathrm{e}}\)-dependence of the scaling of the Gibbs ensemble preparation in \sref{sse:gge_result}.
            %The vertical axis shows the optimized power \(b\) obtained by the fit.        
    \end{figure}

    Here, we discuss the relationship between the energy shell width  \(\varepsilon_{\mathrm{e}}\)  and the scaling behavior of the accuracy of the prepared state.
    Recall that the distance function in Eq.~\eqref{eq:distance} of the main text is given as 
    \begin{eqnarray}
        D(\rho, \rho') = \frac{\|\rho - \rho'\|_{\rm F}}{\sqrt{\|\rho\|_{\rm F}^{2}  - \|\rho'\|_{\rm F}^{2}}},
    \end{eqnarray}
    where \(\|A\|_{\rm F}=\sqrt{{\rm Tr}[A^\dag A]}\) denotes the Frobenius norm.
    In Fig.~\ref{fig:shell_dep}, we show how the scaling exponent \(b\) defined by \(\bar{D}(\rho_{\rm Gibbs}, \rho_t) = O(d^{-b})\) varies according to \(\varepsilon_{\mathrm{e}}\).
    At \(\varepsilon_{\mathrm{e}}=0.1\) for \(L=10\), the corresponding energy shell only includes a single eigenstate. Meanwhile, \(\varepsilon_{\mathrm{e}}=0.75\) corresponds to the extreme case where almost all eigenstates below \(\braket{\hat{H}_{\rm Ising}}_\beta\) are included in the energy shell \([\braket{\hat{H}_{\rm Ising}}_\beta - \varepsilon_{\mathrm{e}} L, \braket{\hat{H}_{\rm Ising}}_\beta]\).
    In Sec.~\ref{sse:ge_result}, we choose intermediate \(\varepsilon_{\mathrm{e}}\) so that the power \(b\) is stable against the choice of \(\varepsilon_{\mathrm{e}}\).

        %We consider \(\varepsilon_{\mathrm{e}}\)-dependence of the scaling of the Gibbs ensemble preparation in \sref{sse:gge_result}.
        %We show the optimized power \(b\) obtained by the fit in \fref{fig:shell_dep}.
        %In \sref{sse:gge_result}, we show the fitted lines and the powers of \(\varepsilon_{\mathrm{e}}=0.5\).
        %For \(\varepsilon_{\mathrm{e}}=0.1\), the corresponding energy shell of \(L=10\) includes only three eigenstates. 
        %For \(\varepsilon_{\mathrm{e}}=0.75\), almost all eigenstates under the corresponding energy \(\braket{\ham_{\mathrm{Ising}}}_{\beta}\) belong to the energy shell \([\braket{\ham_{\mathrm{Ising}}}_{\beta}-\varepsilon_{\mathrm{e}}L,\braket{\ham_{\mathrm{Ising}}}_{\beta}]\).
        %Note that even the ground states are included in the shell for \(L=10\). 
        %In \sref{sse:gge_result}, we choose \(\varepsilon_{\mathrm{e}}\) in the intermediate region of \(\varepsilon_{\mathrm{e}}\) shown in \fref{fig:shell_dep}, where the powers \(b\) do not change significantly.

        %\printfigures
\section{Choosing Lagrange multipliers for the target GGE}\label{app:lagrange}
    The Lagrange multipliers for the GGE in Sec.~\ref{sss:gge} is chosen so that the expectation values of local conserved quantities partly reproduce those of the Gibbs ensemble, while some deviate from it.
    Specifically, we impose the following equalities:
    \begin{align}
            &\tr\left\lbrack \hat{I}_{n}^{+} \rho_{\mathrm{GGE, target}} \right\rbrack =\braket{\hat{I}_{n}^{+}}_{\beta} \ \ \ \,\text{for \(n=0,~1\)},\\
            &\tr\left\lbrack \hat{I}_{n}^{+} \rho_{\mathrm{GGE, target}} \right\rbrack =\braket{\hat{I}_{n}^{+}}_{\beta}+\varepsilon_{\mathrm{I}} L \ \text{for \(2\leq n<\frac{L}{2}\)},\\
            &\tr\left\lbrack \hat{I}_{n}^{-} \rho_{\mathrm{GGE, target}} \right\rbrack =\braket{\hat{I}_{n}^{-}}_{\beta},
    \end{align}
    where \({\braket{\hat{I}_{n}^{\sigma}}_{\beta}}\) represents the expectation values of the LIOMs of the Gibbs ensemble at the inverse temperature \(\beta = 0.4\).
    Note that
    \(\varepsilon_{\mathrm{I}}\) determines the deviation between the target GGE and the Gibbs ensemble.
    For simplicity, we consistently take \(\varepsilon_{\rm I} = 0.05\) for every \(L\).
    In addition, we set \(\tr\left\lbrack \hat{I}_{L/2}^{+} \rho_{\mathrm{GGE, target}} \right\rbrack=0\) because the fermionic tight-binding Hamiltonian obtained from the Jordan-Wigner transformation  (\eref{eq:tight-bind} in the main text) is anti-periodic when the fermionic particle number \(N_{\mathrm{f}}\) is even.
    %We note that we impose a periodic boundary condition on \eref{eq:XX} regarding the Pauli operators.
    
    We remark that the LIOMs with \(n=0,1\) correspond to the total particle number and the energy, respectively. 
    Thus, this target GGE shares only the expectation values of the total particle number and the energy with the Gibbs ensemble.
    %This definition of the target ensures that the target GGE relates to the athermal eigenstate defined in \sref{ss2:gge} for \({\{\hat{I}_{n}^{+}\}}_{n}\) with \(n\geq 2\). 
    Therefore, in order to prepare the subsystem whose size is larger than 2, we need to control additional LIOMs other than the particle number and energy.
    %Note that \(\|\hat{I}_{n}^{\sigma}\|=O(L)\). 

\section{The distance between the GGE and the tGGE}\label{app:tgge_gge_distance}
    \begin{figure}[tbp]
        \centering
        \includegraphics[width=3.4in]{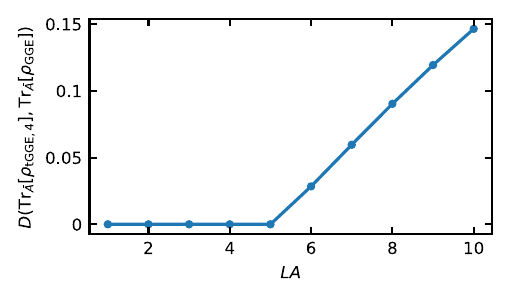}
        \caption{
            The distance, given by \eref{eq:distance}, between  the target GGE in \sref{sse:gge_result} and the corresponding tGGE with \(n_{\mathrm{local}}=4\).
        }
        \label{fig:tGGEvsGGE}
    \end{figure}
    %In \fref{fig:tGGEvsGGE}, we show the distance, given by \eref{eq:distance}, between the tGGE with \(n_{\mathrm{local}}=4\) and the target GGE in \sref{sse:gge_result}.
    To quantify the difference between the tGGE and the target GGE considered in \sref{sse:gge_result}, we show the distance \(D(\rho_{\rm GGE}, \rho_{\rm tGGE})\) in Fig.~\ref{fig:tGGEvsGGE} (see \eref{eq:distance} for the definition).
    We observe that the tGGE and the GGE agree well when \(LA \leq n_{\mathrm{local}}+1\), whereas they deviate when \(LA > n_{\mathrm{local}}+1\).
    This result is compatible with \ccite{fagotti_reduced_2013}, which considers the integrable parameter region of the transverse-field Ising chain.

\section{System-size dependence of learning progress}
    \begin{figure}[tb]
        \centering
        \includegraphics[width=3.4in]{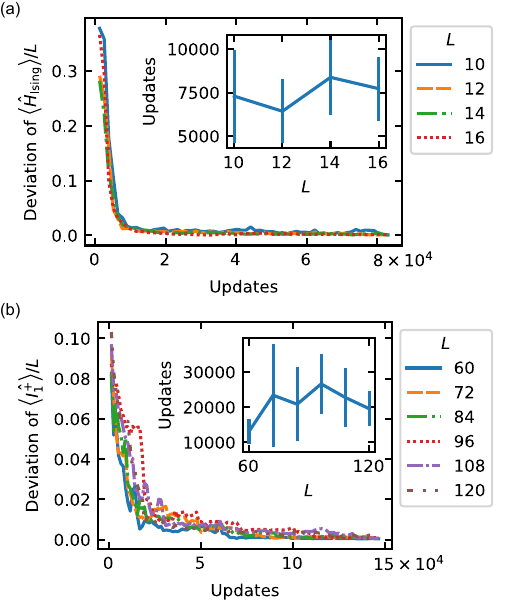}
        \caption{
        The system-size dependence of the learning progress for the local preparation of (a)~Gibbs ensembles in \sref{sse:ge_result} and (b)~GGE in \sref{sse:gge_result}.
        These panels show the deviation of (a)~the energy density and (b)~the LIOM densities, respectively.
        Each panel represents the average regarding different protocols learned independently to perform the finite-size scaling in \fref{fig:nonint_scaling} or \fref{fig:int_scaling}.
        The insets show the number of updates required to bring the deviations below a certain threshold of (a)~\(0.02\) and (b)~\(0.01\), respectively.
        }
        \label{fig:Nupdate_scaling}
    \end{figure}
    In this section, we analyze the system-size dependence of the progresses of RL.
    Figure~\ref{fig:Nupdate_scaling} shows the learning curves of physical observables for different system sizes.
    They tell us that the number of updates required to learn the optimal protocols is almost independent of the system size.
    We suppose that this feature is related to the fact that our method considers only local properties, which are independent of the system size.
    % and can be useful in considering experimental simulations.
    % Therefore, when considering experimental simulation, the computational cost is independent of system size.

\section{Learning for other initial states and unitaries}
    \label{app:add_RL}
    \begin{table}[tb]
        \centering
        \begin{tabular}{lcc}\toprule
            &Initial state&Generators\\\midrule
            (i)&Ground &Same as \sref{sse:ge_result}\vspace{1.5mm}\\
            (ii)&Ground &\(\ham_{\mathrm{Ising}}, \,\sum_{l=1}^LJ\s_{l}^{z}\s_{l+1}^{z}+h\s_{l}^{z},\, g\sum_{l=1}^L\s_{l}^{x}\)\vspace{1.5mm}\\
            (iii)&Product&\(\ham_{\mathrm{Ising}},\, \sum_{l=1}^LJ\s_{l}^{z}\s_{l+1}^{z}+h\s_{l}^{z},\, g\sum_{l=1}^L\s_{l}^{x}\)\vspace{1.5mm}\\
            (iv)&Product&\(\sum_{l=1}^LJ\s_{l}^{z}\s_{l+1}^{z}+h\s_{l}^{z}, \,g\sum_{l=1}^L\s_{l}^{x}\)\vspace{1.5mm}\\
            (v)&Product&\(\sum_{l=1}^L\s_{l}^{z}\s_{l+1}^{z},\,\sum_{l=1}^L\s_{l}^{z},\, \sum_{l=1}^L\s_{l}^{x}\)\\\bottomrule
        \end{tabular}
        \caption{
            The choices of the initial state and the  unitary generators in \aref{app:add_RL}.
            \emph{Ground} means the corresponding initial state of the preparation is the ground state.
            \emph{Product} means that the corresponding initial state of the preparation is the product state, which is the same as that used in \sref{sse:ge_result}.
            }\label{tab:choice_additional_RL}
    \end{table}
    \begin{figure}[tb]
        \centering
        \includegraphics[width=3.4in]{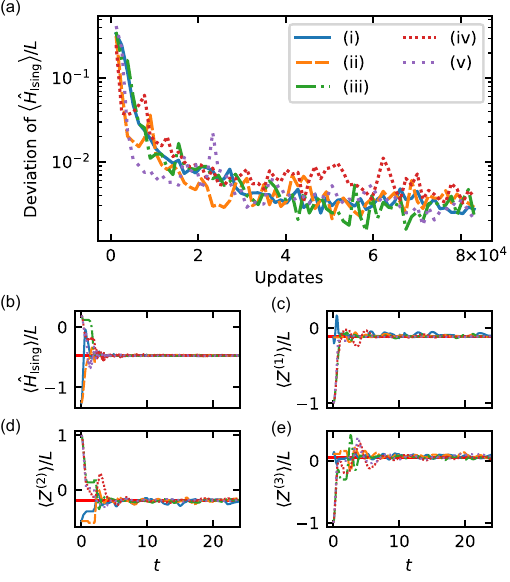}
        \caption{
            The results of deep RL considering different unitary generators and initial states.
            The choices of the unitary generators and the initial state are shown in \tref{tab:choice_additional_RL}.
            (a)~The learning curves for local preparation of the Gibbs ensemble of the transverse field Ising model, which is the same as that used in \sref{sse:ge_result}.
            Each of curves (i)--(v) shows the energy deviation calculated for the prepared states by learning corresponding to each choice.
            The horizontal axis shows the number of parameter updates.
            (b)--(e)~The dynamics of the local observables generated by the state preparation protocol learned by deep RL\@.
            The red horizontal lines show the corresponding expectation values of the target Gibbs ensemble.
            Each panel displays the expectation values as follows: (b)~\(\ham_{\mathrm{Ising}}/L\),
            (c)~\(Z^{(1)}/L=L^{-1}\sum_{l}\s_{l}^{z}\),
            (d)~\(Z^{(2)}/L=L^{-1}\sum_{l}\s_{l}^{z}\s_{l+1}^{z}\), and
            (e)~\(Z^{(3)}/L=L^{-1}\sum_{l}\s_{l}^{z}\s_{l+1}^{z}\s_{l+2}^{z}\).
            The system size \(L=14\), with the inverse temperature set as \(\beta = 0.2\).
        }
        \label{fig:additional_RL}
    \end{figure}
    \begin{figure}[tb]
        \centering
        \includegraphics[width=3.4in]{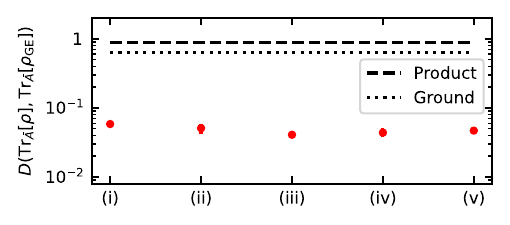}
        \caption{
            The distance between the reduced density operators whose size is 2 of the prepared states and the target Gibbs state used in \sref{sse:ge_result}.
            The red circles represent the prepared state, taking into consideration the different unitary generators and initial states.
            The choices of the unitary generators and the initial state are shown in \tref{tab:choice_additional_RL}.
            The horizontal black dashed line shows the result for the product state which is the same as that used in \sref{sse:ge_result}.
            The horizontal black dotted line represents the ground state.
            The system size \(L=14\).
        }
        \label{fig:additional_RL2}
    \end{figure}
    In this section, we provide the results of deep RL considering different unitary generators and initial states.
    The choices of the unitary generators and the initial state are shown in \tref{tab:choice_additional_RL}.
    The target state is the same as in \sref{sse:ge_result} and the system size \(L=14\).

    Figure~\ref{fig:additional_RL}~(a) shows the learning curves of the RL agent corresponding to choices (i)--(v), respectively.
    % The averaging corresponds to independent learnig with random seeds.
    We can see that as the number of training episodes increases, all RL agents corresponding to the different choices learn the better protocols that achieve smaller energy deviation.

    Figures~\ref{fig:additional_RL} (b)--(e) display the dynamics of the local observables obtained by the preparation protocol learned by the RL agent.
    All of them converge to the corresponding values of the Gibbs ensemble represented by the red horizontal lines.
    
    In \fref{fig:additional_RL2}, we show the time average of the distance function given in \eref{eq:distance} of the subsystems between the target Gibbs state and the prepared state.
    The values in these results are comparable to those in \sref{sse:ge_result} and we can conclude that the prepared states under choices (i)--(v) are typical.

    Surprisingly, even for choices~(iv) and (v), which cannot use \(\ham_{\mathrm{Ising}}\), the observables~(d) and (e), which are not used for the reward, converge to the target values and the distances between the subsystems become small.
    We suppose that the unitary time evolutions, which maintain a steady state with respect to the observables added to the reward~(the energy and total magnetization), are effectively equivalent to the time evolutions by \(\ham_{\mathrm{Ising}}\), which makes the prepared state typical.
    We point out the connection to the studies~\cite{bairey_learning_2019} where the steady state has embedded Hamiltonian information that can be used to infer the parameters of the Hamiltonian.
    Of course, this phenomenon may be model-dependent and needs to be verified more precisely.

    % \renewcommand{\arraystretch}{1}

%\section{The details of the target of the GGE preparation}\label{app:detail_GGE}
\section{Eigenstate thermalization hypothesis}
    \begin{figure}[t]
        \centering
        \includegraphics[width=3.4in]{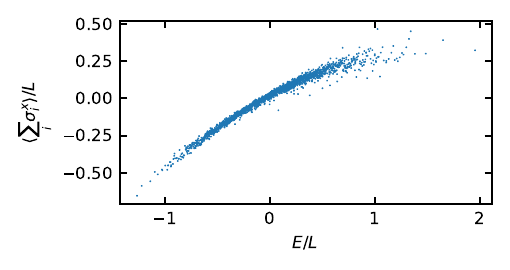}
        \caption{
            The eigenenergy density dependence of the eigenstate expectation values of \(L^{-1}\sum_{l=0}^{L-1}\sigma_{l}^{x}\), 
            that is, we plot \(\{ \braket{E_{\alpha}|\sum_{l=0}^{L-1}\sigma_{l}^{x}|E_{\alpha}}/L \}_{\alpha}\), where \(\left\{ \ket{E_{\alpha}} \right\}_{\alpha}\) are the eigenstates.
            The Hamiltonian is \(\ham_{\mathrm{Ising}}\), whose parameters are the same as the ones used in \sref{sse:ge_result}.
            We can observe that eigenstates in an energy shell share a typical value, which is a microcanonical ensemble average.
            Furthermore, these typical values show the nonlinear dependence on the eigenenergy density.
            The system size \(L=16\).
        }
        \label{fig:nonint_exp_Z_L16}
    \end{figure}
    In this section, we briefly describe the eigenstate thermalization hypothesis (ETH), which is about the typical behavior of eigenstates in an energy shell.
    Furthermore, by looking at the dependence of the eigenstate expectation values of the total magnetization \(\sum_l \sigma_l^z/L\) on the eigenstate energy density, we can infer the cause of the successful local preparation by using the total magnetization as an additional local observable of the reward for RL in \sref{sse:ge_result}.
    
    The ETH refers to the idea that the energy eigenstates satisfy the canonical typicality~\cite{deutsch_quantum_1991,srednicki_chaos_1994,neumann_beweis_1929}.
    The ETH claims that every energy eigenstate in the energy shell represents thermal equilibrium.
    More specifically, the energy eigenstates give the same expectation values of macroscopic observables as the relevant microcanonical ensemble for a large system:
    \begin{align}
        \braket{E_{\alpha}|\hO|E_{\alpha}} \approx \braket{\hO}_{\mathrm{MC}}
    \end{align}
    for every energy eigenstate \(\ket{E_{\alpha}}\) in an energy shell, where \(\Braket{\cdot}_{\mathrm{MC}}\) is the corresponding microcanonical ensemble average.
    Based on the ETH, we can explain the thermalization mechanism of isolated quantum many-body systems~\cite{rigol_thermalization_2008,dalessio_quantum_2016,mori_thermalization_2018}.
    The ETH has been verified numerically for few-body observables in a variety of non-integrable quantum many-body lattice models~\cite{kim_testing_2014,beugeling_finite-size_2014,yoshizawa_numerical_2018,steinigeweg_eigenstate_2013,sorg_relaxation_2014,khodja_relevance_2015,mondaini_eigenstate_2016}.
    
    In addition, a power-law decay with the dimension of the corresponding energy shell is observed for the variance of the energy eigenstate expectation values in the energy shell:
    \begin{align}
        \sigma_{\mathrm{ETH}}^{2}\coloneqq\frac{1}{d}\sum_{\alpha}{\left[ \braket{E_{\alpha}|\hO|E_{\alpha}}-\braket{\hO}_{\mathrm{MC}} \right]}^{2},\label{eq:eth_var}
    \end{align}
    where \(d\) is the dimension of the energy shell,
    that is, the variance decays exponentially with the system size~\cite{beugeling_finite-size_2014}.
    % This result also implies the exponential growth of the accuracy of the preparation utilizing the typicality.

    In \fref{fig:nonint_exp_Z_L16}, we show the eigenenergy density dependence of the eigenstate expectation values of total magnetization \(L^{-1}\sum_{l=0}^{L-1}\sigma_{l}^{x}\), which is the additional local observable of the reward for RL in \sref{sse:ge_result}.
    The Hamiltonian is \(\ham_{\mathrm{Ising}}\), whose parameters are the same as those used in \sref{sse:ge_result}.
    Figure~\ref{fig:nonint_exp_Z_L16} shows that the eigenstates in an energy shell have a typical value. This behavior is consistent with the ETH\@.

    As we note in \sref{ss2:describing_local_preparation_GE}, it is nontrivial to determine how many observables we need to embed the prepared state into a single energy shell.
    In \fref{fig:nonint_exp_Z_L16}, we also observe the nonlinear dependence of the typical values on the energy densities.
    In particular, the dependence appears to be strictly convex.
    % We expect that this dependence helps the prepared state
    We conjecture that this strict convexity helps the prepared state consist of eigenstates within a single energy shell for our demonstration in the non-integrable transverse-field Ising model.

    \section{Numerical methods}
    \subsection{Non-integrable systems}
        For the numerical calculations regarding non-integrable systems, we adopt rigorous standard methods considering the Hilbert space the dimension \(d\) of which scales exponentially with the system size.
        To calculate efficiently, the Hilbert space is resolved by the parity and momentum symmetry.
        Specifically, the calculation is limited to the zero-momentum sector and the parity-symmetric sector.
        The time evolution is performed by simply calculating \(U\ket{\psi}\), which involves the multiplication of a \(d \times d\) matrix by a \(d\)-dimensional vector.
        The construction of the symmetry-resolved basis and the time-evolution are implemented using QuSpin~\cite{weinberg_quspin_2017,weinberg_quspin_2019}.

    \subsection{Non-interacting integrable systems}
        In contrast to the numerical calculation regarding non-integrable systems,
        the calculations regarding integrable systems are done by exploiting the fact that the XX model can be mapped to a free fermionic system.
        In this section, we provide details of the numerical methods used in \sref{sse:gge_result}, which correspond to the preparations in the XX model.
        
        Specifically, we will first discuss the Slater determinant, which efficiently describes free fermionic states, and then the time evolution of the Slater determinant. 
        Next, we explain how to calculate the expectation values of the fermionic observables, and finally how to calculate the expectation values of the observables consisting of hard-core bosons (HCBs), which are used to calculate the observables consisting of Pauli operators.

        \subsubsection{Slater determinant}
            The wave function \(\ket{\psi_{\mathrm{F}}}\) of a free-fermionic system can be represented by a Slater determinant, namely a product of single-particle states:
            \begin{align}
                \ket{\psi_{\mathrm{F}}}=\prod_{m=1}^{N_{\mathrm{f}}}\left( \sum_{l=1}^{L}P_{l,m}\ha_{l}^{\dagger} \right)\ket{0},
            \end{align}
            where \(P\) is the \(L\times N_{\mathrm{f}}\) matrix of components of \(\ket{\psi_{\mathrm{F}}}\) and \(\ket{0}\) represents a vacuum.

        \subsubsection{Time evolution}
            % time-evolution
            The time-evolution of \(\ket{\psi_{\mathrm{F}}}\) under the unitary operator \(\hat{U}\) generated by a quadratic Hamiltonian \(\ham_{\mathrm{q}}:=\sum_{m,n=1}^{L}H_{m,n}\ha_{m}^{\dagger}\ha_{n}\) with time length  \(\delta t\) can be calculated as follows:
            \begin{align}
                \hat{U}\ket{\psi_{\mathrm{F}}}=\prod_{m=1}^{N_{\mathrm{f}}}\left( \sum_{l=1}^{L}\left( UP \right)_{l,m}\ha_{l}^{\dagger} \right)\ket{0}.
            \end{align}
            This calculation is performed by multiplication of an \(L\times L\) unitary matrix \(U=\exp \left[ -iH\delta t \right]\) by an \(L\times N_{\mathrm{f}}\) matrix \(P\).

        \subsubsection{Fermionic observables}
            % fermion
            Consider observables that are quadratic in fermions: \(\hat{A}:=\sum_{m,n=1}^{L}A_{mn}\ha_{m}^{\dagger}\ha_{n}\).
            The expectation values of such observables are calculated as follows:
            \begin{align}
                \braket{\psi_{\mathrm{F}}|\hat{A}|\psi_{\mathrm{F}}}&=\sum_{m,n}A_{mn}\braket{\psi_{\mathrm{F}}|\ha_{m}^{\dagger}\ha_{n}|\psi_{\mathrm{F}}}\\
                &=\sum_{m}A_{mm}-\sum_{m,n}A_{mn}\braket{\psi_{\mathrm{F}}|\ha_{n}\ha_{m}^{\dagger}|\psi_{\mathrm{F}}}\\
                &=\sum_{m}A_{mm}-\sum_{m,n}A_{mn}G_{nm}^{\mathrm{F}},
            \end{align}
            where \(G_{nm}^{\mathrm{F}}:=\braket{\psi_{\mathrm{F}}|\ha_{n}\ha_{m}^{\dagger}|\psi_{\mathrm{F}}}\) is the equal-time Green's function for fermions.

            The creation of a particle at site \(m\) by the action of \(\ha_{m}^{\dagger}\) on \(\ket{\psi_{\mathrm{F}}}\) is represented by the addition of one column to \(P\) with the \(m\)-th element \(P_{m (N_{\mathrm{f}}+1)}=1\) and the rest are \(0\).
            % The \(m\)-th element of the new column is \(1\), and the rest are \(0\).
            In what follows, we denote the new component matrix of the Slater determinant by \(P^{\mathrm{F}(m)}\), which is an \(L\times (N_{\mathrm{f}}+1)\) matrix and is generated by creating a fermion at site \(m\) on the Slater determinant represented by \(P\).
            Because the inner product of two Slater determinants is calculated by the determinant of the product of the component matrices, the equal-time Green's function for fermions is calculated as follows:
            \begin{align}
                G_{nm}^{\mathrm{F}}=\det \left[ \left( P^{\mathrm{F}(n)} \right)^{\dagger}P^{\mathrm{F}(m)} \right].
            \end{align}

            When the columns of \(P\) are orthonormal vectors, we can derive \(\det \left[ \left( P^{\mathrm{F}(n)} \right)^{\dagger}P^{\mathrm{F}(m)} \right]=\delta_{nm}-\sum_{k=1}^{N_{\mathrm{f}}}P_{nk}P_{mk}^{*}\), which results in
            \begin{align}
                \braket{\psi_{\mathrm{F}}|\hat{A}|\psi_{\mathrm{F}}}=\tr\left[ P^{\dagger} AP \right].
            \end{align}
        
        \subsubsection{Hard-core bosonic observables}
            % boson
            Next, we consider how to compute the expectation values of observables consisting of Pauli operators.
            In this section, we consider HCBs in order to introduce the creation and annihilation picture of particles.
            Here, we denote the creation and annihilation operators for a HCB acting on site \(m\) by \(\hb_{m}^{\dagger}\) and \(\hb_{m}\), respectively.
            The HCB operators are introduced as \(\hb_{m}^{\dagger}=\left( \s_{m}^{x}+i\s_{m}^{y} \right)/2,\,\hb_{m}=\left( \s_{m}^{x}-i\s_{m}^{y} \right)/2\).
            % Next, we consider the calculation of the expectation values of HCB operators, which is used for the calculation regarding Pauli operators.
            % Therefore, HCB operators have the relation like .
            The calculating method described here follows the technique used in \cscite{rigol_relaxation_2007,rigol_emergence_2004,rigol_universal_2004,rigol_free_2005}.
            
            Consider observables that are quadratic in HCBs: \(\hat{B}:=\sum_{m,n=1}^{L}B_{mn}\hb_{m}^{\dagger}\hb_{n}\).
            The expectation values of such observables are calculated as follows:
            \begin{align}
                \braket{\psi_{\mathrm{F}}|\hat{B}|\psi_{\mathrm{F}}}=\sum_{m}B_{mm}\left( 1- G_{mm}^{\mathrm{B}}\right)+\sum_{m\neq n}B_{mn}G_{nm}^{\mathrm{B}},
            \end{align}
            where \(G^{\mathrm{B}}_{nm}:=\braket{\psi_{\mathrm{F}}|\hb_{n}\hb_{m}^{\dagger}|\psi_{\mathrm{F}}}\) is the equal-time Green's function for HCBs.

            The action of \(\hb_{m}^{\dagger}=\ha_{m}^{\dagger}\prod_{l=1}^{m-1}e^{-i\pi \ha_{l}^{\dagger}\ha_{l}}\) on \(\ket{\psi_{\mathrm{F}}}\) is represented by a change of sign on the element \(P_{kl}\) for \(k\leq m-1\) 
            and then the addition of one column to \(P\), where the \(m\)-th element \(P_{m (N_{\mathrm{f}}+1)}=1\) and the rest are \(0\).
            % The \(m\)-th element of the new column is \(1\) and the rest are \(0\). 
            As a result, the Green's function for HCBs is calculated as follows:
            \begin{align}
                G^{\mathrm{B}}_{nm}=\det\left[ \left( P^{\mathrm{B}(n)} \right)^{\dagger}P^{\mathrm{B}(m)} \right],
            \end{align}
            where \(P^{\mathrm{B}(m)}\) is the new component matrix of the Slater determinant, which is generated by creating a HCB at site \(m\) on the Slater determinant represented by \(P\).

%\bibliographystyle{apsrev4-2}
% \bibliography{library}
    
\end{document}